\documentclass[11pt]{article}
\usepackage{latexsym,amssymb,amsmath}
\usepackage[dvips]{graphics,graphicx,epsfig,color}
\usepackage[lofdepth,lotdepth]{subfig}
\begin{document}
\title{\bf Resonant Seismic Motion Within a Sedimentary Basin Revisited}
\author{Armand Wirgin\thanks{retired from LMA, CNRS, UPR 7051, Aix-Marseille Univ, Centrale Marseille, F-13453 Marseille Cedex 13, France, ({\tt armand.wirgin@gmail.com})} }
\date{\today}
\maketitle
\begin{abstract}
The problem, of 2D canonical nature, examined herein in the space-frequency framework,  concerns a SH-polarized plane body seismic wave propagating in a hard, non lossy half space (bedrock) containing  a soft, lossy cylindrical basin of semi-circular shape. The displacement response, at points both within and outside the  basin is represented alternatively in integral (for basins of general shape) and partial wave (for basins of semi-circular shape) forms, the integral representation leading to an expression of a sort of conservation of energy relation, and an explanation of how the energetic gain provided by the incident wave, is divided essentially into two sources of loss: radiation damping (taking place in the bedrock) and volumic material damping (taking place within the basin), whereas the partial wave representation enables a detailed theoretical demonstration of the resonant nature of the response, marked by amplification and concentration of the displacement field essentially within the basin at certain (eigen-)frequencies and very little, or no amplification and concentration, at other frequencies. These phenomena  carry over to the volumic material damping and radiation damping loss functions (of frequency), with the equivalent of resonant displacement response residing in peaks of the volumic material damping loss function and troughs of the radiation damping loss function. Maps, for the semi-circular basin, are computed herein of the displacement field both at resonant and non-resonant frequencies which strongly evoke those (of numerical nature resulting from the boundary element scheme) appearing in two publications by other authors for basins of more realistic shape, so as to suggest that the results in the present study are applicable, for a large part, to basins of general shape.
\end{abstract}
\newline
\newline
Abbreviated title: Concentrated seismic motion in basins
\newline
\newline
Corresponding author: Armand Wirgin, \\ e-mail: armand.wirgin@gmail.com
\newpage
\tableofcontents
\newpage
\newpage
\section{Introduction}\label{intro}
Populated areas (i.e., settlements in the formulation of \cite{tr71}) are frequently  underlain by layers, valleys or canyons (in short: basins) filled partially or fully with sediments. The response to a seismic disturbance has often been observed to be stronger in such settlements than in structurally-similar settlements built on hard rock sites \cite{kb81,tk84}.

At first, it was thought that the reason for this response-enhancement is that the basin (surrounded by hard rock except at ground level) acts like a sort of lens \cite{ja71} which focuses the incident seismic wave to the ground at locations occupied by the settlement. A flat layer-of-infinite-lateral-extent(FLOILE)(also frequently-termed 1D) model of the basin \cite{wi16} attributed the amplified ground motion to interference phenomena, often qualified as 'resonances' (by the fact that the ground response appears to be at its maximum at a series of well-defined frequencies)\cite{wi16}. The FLOILE model then led to various models of layers of finite lateral extent \cite{tr71,bb80,mr95,kt14,wi16} in order to account for what appears to be enhanced seismic motion on the ground near the corners of the basin \cite{ka96,sl25}, thought to be the locations at which are generated the motion-amplification agents called 'surface waves' \cite{bb80}. A rigorous analysis \cite{tr71} (followed by \cite{tl91,wi95,wd95,kt14,pm21}) of these (not-all) phenomena appealed to a semi-circular cylindrical totally-filled basin model whereby the motion was shown to be concentrated and amplified essentially on the portion of ground above the basin and not elsewhere (i.e., not on its flanks along the ground).

Many of these (and other) studies (see, for instance: \cite{mp71,wt74,le84,tl91,fa95,wd95,kl03,lz04,sk05,sk08,pm21,cp25}) employ terms such as 'response aggravation' \cite{rm16},  'amplification', as well as the (the above-mentioned) concepts of 'resonances', and 'surface waves' \cite{bb80,ka96,sl25}, but do not clearly establish the connection between these concepts and terms. Consequently, the purpose of the present contribution is to contribute to filling in this explanatory gap, as concerns the response on the ground and in  the underground (this latter  aspect could be useful for those who are interested in the seismic response of tunnels, cavities \cite{lz04} and other buried structures)  to a seismic disturbance in a hard half space containing a basin in the absence of the above-ground settlement.

I shall revisit herein (see \cite{wi95,wi16} for my earlier contributions to the subject of) the semi-circular cylindrical totally-filled sedimentary basin model  and mathematically obtain an expression for something akin to a conservation of energy relation. The latter shows that the majority of the incident energy (more precisely 'flux') is redistributed into absorbed energy (flux) within the basin and re-radiated energy (flux) into the bedrock, the small remainder being associated with energy (flux) exchanges across the lower interface of the basin. The absorbed energy (flux) is, by nature, confined to the entire space occupied by the basin, and shown to increase substantially (this being related to the previously-mentioned amplification and aggravation) at, or near, the {\it resonance frequencies} (their origin being explained) of the basin  while the re-radiated energy (flux) in the bedrock is relatively-small or nil. At these frequencies, the motion (as represented by the displacement) is shown to be concentrated and relatively large at a series of hot spots within the basin (and, by extension, on the portion of ground constituting the upper boundary of the basin).

The question of how, and to what extent, a part of the incident energy  (flux) is transferred to the above-ground built component of the settlement (called the soil-structure interaction problem \cite{cc02,kw21}) will be addressed in a subsequent study.
\section{The boundary  value problem}\label{bvp}
The ground, occupying $y=0$ (in the $xyz$ cartesian coordinate system), is flat and separates two half-spaces. The above-ground half-space  region ($y>0$) is filled with air, assumed to be the vacumn, within which no elastic wave can propagate. The underground half-space region ($y<0$) is filled with a hard material $M^{[0]}$ except within a semi-circular cylindrical basin (center at $x=y=0$) whose upper, flat boundary is at ground level and lower boundary described by $x^2+y^2=h^2; \forall z\in \mathbb{R}$. This basin is  assumed to be entirely filled with the soft, lossy material $M^{[1]}$. The two materials are assumed to be in welded contact at the lower boundary of the basin.

Since the propagation vector of the incident SH body wave lies in the $x-y$ plane and the basin boundaries do not depend on $z$, the problem is 2D and lends itself to an analysis in the sole $x-y$ (sagittal) plane. Thus, from now on, everything refers to what exists and occurs in this sagittal plane. Finally, I carry out the analysis in the space-frequency framework, so that the displacement field $\mathbf{U}$ depends on time  $t$ via $\mathbf{U}=\mathbf{u}\exp(-i\omega t)$, with $\mathbf{u}$ the vectorial displacement function of interest herein. The latter, as well as the stress fields, and constitutive parameters depend on the frequency $f$ (related to the angular frequency $\omega$ by $\omega=2\pi f$). This dependence will henceforth be implicit. Similarly, the $z$-component of the vectorial displacement field, designated by the scalar $u_{z}(\mathbf{x},f)$, with $\mathbf{x}=(x,y)$, will be written as $u(\mathbf{x})$).

\begin{figure}[ht]
\begin{center}
\includegraphics[width=0.75\textwidth]{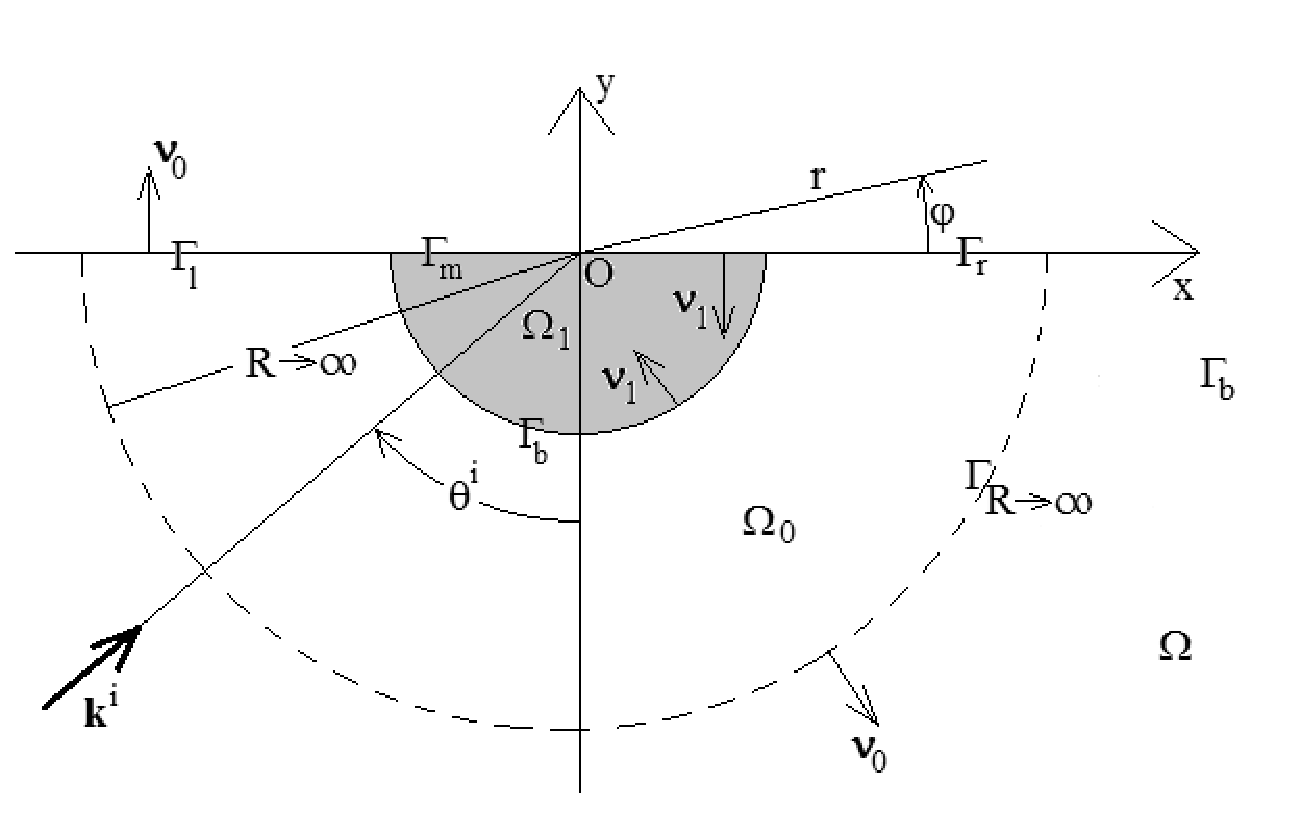}
\caption{Sagittal plane view of 2D scattering configuration. }
\label{softcircbasin}
\end{center}
\end{figure}

Fig. \ref{softcircbasin} describes the scattering configuration. In this figure, $\mathbf{k}^{i}=\mathbf{k}^{i}(\theta^{i},\omega)$ is the incident wavevector oriented so that its $z$  component is nil and $\theta^{i}$  the angle of incidence. The basin occupies the finite-sized region $\Omega_{1}$ and what is exterior to it (although below $y=0$) is the region $\Omega_{0}$. $M^{[0]}$ is a hard (rock-like), lossless, linear, isotropic, homogeneous (LIH)  solid whereas $M^{[1]}$ is a soft (sediment-like), {\it lossy}, LIH solid.

The flat ground, which is the locus of a stress-free boundary condition, is designated by $\Gamma_{g}$ and is composed  of three segments; $\Gamma_{l}$, $\Gamma_{m}$, and $\Gamma_{r}$. The upper and lower boundaries of the basin are $\Gamma_{m}$ and $\Gamma_{b}$ respectively, and $\Gamma_{b}$ is the (curved) locus of a continuity of  displacement condition and a continuity of stress condition.

The medium within $\Omega_{0}$ is characterized by its (real) density $\rho^{[0]}$ and (real) shear body-wave velocity $\beta^{[0]}=\beta^{'[0]}+i0$ with $\beta^{'[0]}\geq 0$, whereas the medium within $\Omega_{1}$ is characterized by its (real) density $\rho^{[0]}\geq 0$ and  and (complex)  body-wave velocity $\beta^{[1]}=\beta^{'[1]}+i\beta^{''[1]}$, with $\beta^{'[1]}\geq 0$ and $\beta^{''[1]}\leq 0$.  The shear modulii derive from the densities and velocities via $\mu=\rho\beta^2$ from which it ensues that $\mu^{[0]}=\mu^{'[0]}+i0$, with $\mu^{'[0]}\geq 0$, and $\mu^{[1]}=\mu^{'[1]}+i\mu^{''[1]}$ with $\mu^{'[1]}\geq 0$ and $\mu^{''[1]}\leq 0$. I shall assume that the lossy, dispersive nature of $M^{[1]}$ can be taken into account by the constant-$Q$ relation of Kjartansson    \cite{kj79}
\begin{equation}\label{1-002}
\beta^{[1]}(\omega)=\Re(\beta_{ref}^{[1]})
\left(\frac{-i\omega}{\omega_{ref}}\right)
^{\frac{1}{\pi}tan^{-1}\left(\frac{1}{Q^{[1]}}\right)}~
\end{equation}
with $\omega_{ref}=0.09 rad/s$ \cite{gr05} and $\beta_{ref}^{[1]}$ the $\omega=\omega_{ref}$ value of $\beta^{[1]}$, so that the given constitutive elastic parameters (which are all constants with respect to $f$) of the problem become: $\rho^{[0]}$, $\beta^{[0]}$, $\rho^{[1]}$ and $\beta_{\infty}^{[1]}$ and $Q^{[1]}$. The sole given geometrical parameter is, of course, the radius $h$ of the basin.

The incident plane (body) wave field is of the form
\begin{equation}\label{1-008}
u^{i}(\mathbf{x})=A^{i}\exp(i\mathbf{k}^{i}\cdot\mathbf{x})=A^{i}\exp[i(k_{x}^{i}x+k_{y}^{i}y)]~,
\end{equation}
wherein $A^{i}=A^{i}(\omega)$ is the spectral amplitude of the incident seismic pulse, $\mathbf{k}^{i}=(k_{x}^{i},k_{y}^{i})$, $k_{x}^{i}=k^{[0]}\sin\theta^{i}$, $k_{y}^{i}=k^{[0]}\cos\theta^{i}$, $k^{[l]}=\omega/\beta^{[l]}~;~l=0,1$. $A^{i}$ (assumed herein not to depend on $f$) and $\theta^{i}$ are the two given excitation parameters of the problem.

Owing to the fact that the configuration comprises two distinct regions, each in which the elastic parameters are constants as a function of the space variables, it is opportune to employ domain decomposition (DC) and (later on) separation of variables (SOV). Thus, I decompose the total  field $u$ as:
\begin{equation}\label{1-010}
u(\mathbf{x})=u^{[j]}(\mathbf{x})~;~\forall\mathbf{x}\in\Omega_{j},~j=0,1~,
\end{equation}
with the understanding that these fields satisfy the 2D SH frequency domain elastic wave equation (i.e., Helmholtz equation)
\begin{equation}\label{1-020}
\Big(\triangle+\big(k^{[j]}\big)^{2}\Big)u^{[j]}(\mathbf{x})=0~;~\forall\mathbf{x}\in\Omega_{j},~j=0,1~,
\end{equation}
 with the notations $\triangle=\frac{\partial^{2}}{\partial x^{2}}+\frac{\partial^{2}}{\partial y^{2}}$ in the cartesian coordinate system of the sagittal plane.

 In addition, the field $u^{[0]}$ satisfies the radiation condition
\begin{equation}\label{1-030}
u^{[0]}(\mathbf{x})-u^{i}(\mathbf{x})\sim \text {outgoing~wave} ~;~\|\mathbf{x}\|\rightarrow \infty~.
\end{equation}
due to the fact   that $\Omega_{0}$, which  corresponds to taking $\lim R\rightarrow \infty$ in fig. \ref{softcircbasin}, is unbounded (i.e., a semi-infinite domain).

The stress-free nature of the boundaries $\Gamma_{g}=\Gamma_{l}\cup\Gamma_{m}\cup\Gamma_{r}$, entails the boundary conditions:
\begin{equation}\label{bc-010}
\mu^{[0]}\boldsymbol{\nu}_{0}\cdot\nabla u^{[0]}(\mathbf{x})=0~;~\forall\mathbf{x}\in\Gamma_{l}\cup\Gamma_{r}~,
\end{equation}
\begin{equation}\label{bc-020}
\mu^{[1]}\boldsymbol{\nu}_{1}\cdot\nabla u^{[1]}(\mathbf{x})=0~;~\forall\mathbf{x}\in\Gamma_{m}~,
\end{equation}
wherein   wherein $\boldsymbol{\nu}_{j}$ is the unit vector normal to the boundary $\partial{\Omega_{j}}$ of $\Omega_{j}$ and $\nabla$ designates the gradient operator.

Finally, the fact that $\Gamma_{b}$ was assumed to be an interface across which two media are in welded contact, entails the continuity conditions:
\begin{equation}\label{bc-040}
u^{[0]}(\mathbf{x})-u^{[1]}(\mathbf{x})=0~;~\forall\mathbf{x}\in\Gamma_{b}~,
\end{equation}
\begin{equation}\label{bc-050}
\mu^{[0]}\boldsymbol{\nu}_{1}\cdot\nabla u^{[0]}(\mathbf{x})-\mu^{[1]}\boldsymbol{\nu}_{1}\cdot\nabla u^{[1]}(\mathbf{x})=0~;~\forall\mathbf{x}\in\Gamma_{b}~.
\end{equation}

The purpose of addressing such a boundary-value problem is  generally to determine $u^{[j]}(\mathbf{x});~j=0,1$ for various solicitations  and parameters  relative to the  geometries of, and media filling, $\Omega_{j}~;~j=0,1$. A specific, additional, feature of the following analysis is the establishment, verification,  and application of an energy-like (termed flux) conservation law. To show that this conservation law makes sense, I shall appeal to the well-known rigorous, closed-form solution of the problem.
\section{Ingredients of the conservation law}\label{icl}
One may ask what the purpose is of establishing a conservation of energy (flux) law for a basin submitted to a seismic load. My opinion is that understanding the energy (flux) exchanges that occur beneath ground level provides the means for understanding the energy (flux) exchanges above ground level in the presence of a settlement comprising a built environment (and therefore the seismic damage potential of the latter), which is at the heart of the soil-structure  interaction problem \cite{cc02,wi16}.

The analysis in this section is applicable to a basin of any (i.e., not necessarily semi-circular) shape in the absence of an above-ground built environment. This means that everything is as in sect.\ref{bvp}  ~except that now $\Gamma_{b}$ {\it is not-necessarily a semi-circular arc}. Furthermore, contrary to what was assumed in \cite{wi19}, herein I consider the shear modulus of the material in the basin  to be {\it complex}).

Eq. (\ref{1-020}) yields
\begin{equation}\label{2-020}
\Big(\triangle+\big[\big(k^{[j]}\big)^{2}\big]^{*}\Big)u^{[j]*}(\mathbf{x})=0~;~\forall\mathbf{x}\in\Omega_{j},~l=0,1~,
\end{equation}
wherein $(X+iY)^{*}=X-iY$. It follows (with $d\varpi$ the differential surface element) that
\begin{equation}\label{2-030}
\int_{\Omega_{j}}u^{[j]*}(\mathbf{x})\Big\{\Big(\triangle+\big(k^{[j]}\big)^{2}\Big)u^{[j]}(\mathbf{x})-
u^{[j]*}(\mathbf{x})\Big(\triangle+\big[\big(k^{[j]}\big)^{2}\big]^{*}\Big)u^{[j]}(\mathbf{x})\Big\}d\varpi
=0~;~j=0,1~,
\end{equation}
or
\begin{multline}\label{2-040}
\int_{\Omega_{j}}\Big\{u^{[j]*}(\mathbf{x})\triangle u^{[j]}(\mathbf{x})-u^{[j]}(\mathbf{x})\triangle u^{*[j]}(\mathbf{x})\Big\}d\varpi+\\
\int_{\Omega_{j}}\Big\{\big(k^{[j]}\big)^{2}-\big[\big(k^{[j]}\big)^{2}\big]^{*}\Big\}|u^{[j]}(\mathbf{x})|^{2}d\varpi
=0~;~j=0,1~.
\end{multline}
 I intend to apply Green's second identity to the first integral, and to do this I must define the boundaries of $\Omega_{j}~;~j=0,1$. I already know that the  (closed) boundary  of $\Omega_{1}$ is constituted by the union of $\Gamma_{m}$ and $\Gamma_{b}$, but until now, $\Omega_{0}$ was not closed. To close it (see fig.\ref{softcircbasin}), I imagine a semicircle $\Gamma_{\mathcal{R}}$, of large radius $\mathcal{R}$ centered at the origin $O$,  to be drawn so as to intersect the ground at $x=\pm \mathcal{R}$ and to intersect the $y$ axis at $y=-\mathcal{R}$. Thus, the closed boundaries of $\Omega_{1}$ and $\Omega_{0}$ are:
\begin{equation}\label{2-050}
\partial_{\Omega_{1}}=\Gamma_{m}\cup\Gamma_{b}~~,~~
\partial_{\Omega_{0}}=\Gamma_{l}\cup\Gamma_{b}\cup\Gamma_{r}\cup\Gamma_{\mathcal{R}\rightarrow\infty}~,
\end{equation}
and, as in fig. \ref{softcircbasin},  $\boldsymbol{\nu}_{0}$ designates the unit vector normal to $\partial\Omega_{0}$  pointing towards the exterior of $\partial\Omega_{0}$ whereas  $\boldsymbol{\nu}_{1}$ designates the unit vector normal to $\partial\Omega_{1}$  pointing towards the interior of $\partial\Omega_{1}$ . I now apply Green's second identity to obtain (with $d\gamma$ the differential arc element)
\begin{multline}\label{2-060}
\int_{\partial\Omega_{0}}\Big\{u^{[j]*}(\mathbf{x})\boldsymbol{\nu}_{0}\cdot\nabla u^{[j]}(\mathbf{x})-u^{[0]}(\mathbf{x}) \boldsymbol{\nu}_{0}\cdot\nabla u^{*[0]}(\mathbf{x})\Big\}d\gamma+\\
\int_{\Omega_{0}}\Big\{\big(k^{[0]}\big)^{2}-\big[\big(k^{[0]}\big)^{2}\big]^{*}\Big\}|u^{[0]}(\mathbf{x})|^{2}d\varpi
=0~,
\end{multline}
\begin{multline}\label{2-065}
-\int_{\partial\Omega_{1}}\Big\{u^{[1]*}(\mathbf{x})\boldsymbol{\nu}_{1}\cdot\nabla u^{[1]}(\mathbf{x})-u^{[1]}(\mathbf{x}) \boldsymbol{\nu}_{1}\cdot\nabla u^{*[1]}(\mathbf{x})\Big\}d\gamma+\\
\int_{\Omega_{1}}\Big\{\big(k^{[1]}\big)^{2}-\big[\big(k^{[1]}\big)^{2}\big]^{*}\Big\}|u^{[1]}(\mathbf{x})|^{2}d\varpi
=0~,
\end{multline}
or, due to the lossless nature of $M^{[0]}$, the homogeneous nature of $M^{[1]}$ and the boundary conditions (\ref{bc-010})-(\ref{bc-020}):
\begin{equation}\label{2-070}
\Im\int_{\partial\Omega_{0}}u^{[0]*}(\mathbf{x})\boldsymbol{\nu}_{0}\cdot\nabla u^{[0]}(\mathbf{x})d\gamma=0~,
\end{equation}
\begin{equation}\label{2-080}
-\Im\int_{\partial\Omega_{1}}u^{[1]*}(\mathbf{x})\boldsymbol{\nu}_{1}\cdot\nabla u^{[1]}(\mathbf{x})d\gamma+\Im\big[\big(k^{[1]}\big)^{2}\big]\int_{\Omega_{1}}|u^{[1]}(\mathbf{x})|^{2}d\varpi=0~.
\end{equation}
More explicitly, and on account of (\ref{2-050}),
\begin{multline}\label{2-090}
\Im\int_{\Gamma_{l}\cup\Gamma_{r}}u^{[0]*}(\mathbf{x})\boldsymbol{\nu}_{0}\cdot\nabla u^{[0]}(\mathbf{x})d\gamma+
\Im\int_{\Gamma_{b}}u^{[0]*}(\mathbf{x})\boldsymbol{\nu}_{0}\cdot\nabla u^{[0]}(\mathbf{x})d\gamma+\\
\Im\int_{\Gamma_{\infty}}u^{[0]*}(\mathbf{x})\boldsymbol{\nu}_{0}\cdot\nabla u^{[0]}(\mathbf{x})d\gamma=0~,
\end{multline}
\begin{equation}\label{2-100}
-\Im\int_{\Gamma_{m}}u^{[1]*}(\mathbf{x})\boldsymbol{\nu}_{1}\cdot\nabla u^{[1]}(\mathbf{x})d\gamma-
\Im\int_{\Gamma_{b}}u^{[1]*}(\mathbf{x})\boldsymbol{\nu}_{1}\cdot\nabla u^{[1]}(\mathbf{x})d\gamma+\Im\big[\big(k^{[1]}\big)^{2}\big]\int_{\Omega_{1}}|u^{[1]}(\mathbf{x})|^{2}d\Omega=0~.
\end{equation}
wherein $\Gamma_{\infty}=\Gamma_{\mathcal{R}\rightarrow\infty}$. Due to the (stress-free) boundary  conditions (\ref{bc-010})-(\ref{bc-020}), I find:
\begin{equation}\label{2-106}
\Im\int_{\Gamma_{b}}u^{[0]*}(\mathbf{x})\boldsymbol{\nu}_{0}\cdot\nabla u^{[0]}(\mathbf{x})d\gamma+\\
\Im\int_{\Gamma_{\infty}}u^{[0]*}(\mathbf{x})\boldsymbol{\nu}_{0}\cdot\nabla u^{[0]}(\mathbf{x})d\gamma=0~,
\end{equation}
\begin{equation}\label{2-108}
-\Im\int_{\Gamma_{b}}u^{[1]*}(\mathbf{x})\boldsymbol{\nu}_{1}\cdot\nabla u^{[1]}(\mathbf{x})d\gamma+\Im\big[\big(k^{[1]}\big)^{2}\big]\int_{\Omega_{1}}|u^{[1]}(\mathbf{x})|^{2}d\Omega=0~.
\end{equation}
It is fairly-easy to verify that each of these two expressions is trivially true (i.e., is independent of the correct {\it solutions} for $u^{[j]}~;~ j=0,1$, provided one employs in each expression the correct {\it representation} of $u^{[j]}~;~ j=0,1$). Another "inconvenience" is that two equations are one too many to express a conservation law which must be expressed by a single equation. To effectuate the necessary reduction I make use of the continuity conditions (\ref{bc-040}) and (\ref{bc-050}) whereby (\ref{2-108}) becomes:
\begin{equation}\label{2-130}
-\Im\int_{\Gamma_{b}}\frac{\mu^{[0]}}{\mu^{[1]}}u^{[0]*}(\mathbf{x})\boldsymbol{\nu}_{1}\cdot\nabla u^{[0]}(\mathbf{x})d\Gamma+
\Im\big[\big(k^{[1]}\big)^{2}\big]\int_{\Omega_{1}}|u^{[1]}(\mathbf{x})|^{2}d\Omega=0~,
\end{equation}
or more explicitly:
\begin{multline}\label{2-140}
-\Re{\Bigg [}\frac{\mu^{[0]}}{\mu^{[1]}}{\Bigg ]}\Im\int_{\Gamma_{b}}u^{[0]*}(\mathbf{x})\boldsymbol{\nu}_{1}\cdot\nabla u^{[0]}(\mathbf{x})d\Gamma-\\
-\Im{\Bigg [}\frac{\mu^{[0]}}{\mu^{[1]}}{\Bigg ]}\Re\int_{\Gamma_{b}}u^{[0]*}(\mathbf{x})\boldsymbol{\nu}_{1}\cdot\nabla u^{[0]}(\mathbf{x})d\Gamma+\Im\big[\big(k^{[1]}\big)^{2}\big]\int_{\Omega_{1}}|u^{[1]}(\mathbf{x})|^{2}d\Omega=0~.
\end{multline}
On account of (\ref{2-106}) I finally find:
\begin{multline}\label{2-145}
\Re{\Bigg [}\frac{\mu^{[0]}}{\mu^{[1]}}{\Bigg ]}\Im\int_{\Gamma_{\infty}}u^{[0]*}(\mathbf{x})\boldsymbol{\nu}_{1}\cdot\nabla u^{[0]}(\mathbf{x})d\gamma-\\
-\Im{\Bigg [}\frac{\mu^{[0]}}{\mu^{[1]}}{\Bigg ]}\Re\int_{\Gamma_{b}}u^{[0]*}(\mathbf{x})\boldsymbol{\nu}_{1}\cdot\nabla u^{[0]}(\mathbf{x})d\Gamma+\Im\big[\big(k^{[1]}\big)^{2}\big]\int_{\Omega_{1}}|u^{[1]}(\mathbf{x})|^{2}d\Omega=0~.
\end{multline}
which is the {\it primitive} form of the sought-for conservation law expression.

Further on, I will show that the first term in this relation comprises the incident and re-radiated energy (flux), the second term energy (flux) exchanges along the lower boundary of the basin, and the third term absorbed energy (flux) {\it within the basin}. Note should be taken of the fact that nowhere in this relation appear the seismic motions {\it at specific points of, or all along, the ground}, which are where qualities such as 'amplification' or 'aggravation' factors are usually measured or computed. Likewise, to my knowledge, 'amplification' and 'aggravation' factors don't add up to a mathematical constant, and are consequently not conserved. This fact makes it difficult to compare the seismic response: of one basin to another (for a given seismic load), or due to one seismic load or another (for a given basin) via the  concepts of 'amplification' or 'aggravation'.
\section{Separation of variables representations of the two displacement fields}
The material in this section is in close relation to that of references \cite{mp71}, \cite{tr71} and \cite{wi95}. It is convenient to rewrite the radiation condition as
\begin{equation}\label{3-010}
u^{s}(\textbf{x})=u^{[0]}(\textbf{x})-u^{i}(\textbf{x})-u^{r}(\textbf{x})\sim \text{outgoing wave} ;   ||\textbf{x}||\rightarrow\infty ,
\end{equation}
wherein $u^{r}(\textbf{x})$ is the so-called reflected {\it outgoing} plane wave ($u^{i}(\textbf{x})$ being {\it incoming} with respect to the basin)
\begin{equation}\label{3-020}
u^{r}(\textbf{x})=A^{i}\exp[i(k_{x}^{i}x-k_{y}^{i}y)]~ ,
\end{equation}
and $u^{s}(\textbf{x})$ is the so-called {scattered} wave (which must be {\it outgoing} by virtue of (\ref{3-010})). Note that both $u^{i}(\textbf{x})$ and $u^{r}(\textbf{x})$, as well as $u^{s}(\textbf{x})$, satisfy the Helmholtz wave equation in $\Omega_{0}$ and the stress-free relation on $\Gamma_{g}$
\begin{equation}\label{3-030}
u^{i}_{,y}(\textbf{x})+u^{r}_{,y}(\textbf{x})=0~;\forall ~{\bf x}\in\Gamma_{g}~ ,
\end{equation}
wherein $F_{,y}=\partial F/\partial y$.
The SOV representation of $u^{s}(\textbf{x})$ in the  $(r,\phi)$ polar coordinate system, which satisfies the Helmholtz equation in $\Omega_{0}$, the stress-free boundary condition (\ref{1-010}) and the radiation condition (\ref{3-030}), is
\begin{equation}\label{3-040}
u^{s}({\bf{x}})=\sum_{n=0}^{\infty}B_{n}H_{n}^{(1)}(k^{[0]}r)\cos(n\phi)~;~\forall \mathbf{x}\in \Omega_{0}~ ,
\end{equation}
wherein $H_{n}^{(1)}(.)$ is the Hankel function of the first kind for which the superscript is henceforth implicit.

 I now address the problem of the conversion of $u^{i}(\textbf{x})+u^{r}(\textbf{x})$ to polar coordinates. This is based on the employment of formula (9.1.41) in \cite{as68} whereby
\begin{equation}\label{3-050}
u^{i}({\bf{x}})=A^{i}\sum_{n=-\infty}^{\infty}J_{n}(k^{[0]}r)\exp[i n(\theta^{i}+\phi)]~,~
u^{r}({\bf{x}})=A^{i}\sum_{n=-\infty}^{\infty}J_{n}(k^{[0]}r)\exp[i n(\theta^{i}-\phi)]~ ,
\end{equation}
in which $J_{n}(.)$ is the Bessel function. It is then easy to obtain:
\begin{equation}\label{3-060}
u^{i}({\bf{x}})+u^{r}({\bf{x}})=\sum_{n=0}^{\infty}A_{n}^{[0]}J_{n}(k^{[0]}r)\cos(n\phi)~,~
\end{equation}
wherein
\begin{equation}\label{3-070}
A_{n}^{[0]}=2A^{i}\epsilon_{n}\exp(in\pi/2)\cos(n(\theta^{i}-\pi/2))~,~
\end{equation}
with $\epsilon_{n}$ the Neumann factor (=1 for $n=0$ and =2 for $n>0$), it being understood that the $A_{n}^{[0]}$ are known a priori since $A^{i}$ and $\theta^{i}$ are given parameters for the problem at hand.

It ensues that the {\it total} displacement field in $\Omega_{0}$ admits to the polar coordinate so-called {\it partial wave} representation
\begin{equation}\label{3-080}
u^{[0]}({\bf{x}})=\sum_{n=0}^{\infty}U_{n}^{[0]}(r)\cos(n\phi)=\sum_{n=0}^{\infty}\left(A_{n}^{[0]}J_{n}(k^{[0]}r)+B_{n}^{[0]}H_{n}(k^{[0]}r)\right)\cos(n\phi)~;~\forall\mathbf{x}\in\Omega_{0}~.~
\end{equation}
I obtain, by the same SOV technique, the following {\it partial wave} (i.e. $U_{n}^{[j]}(r)\cos(n\phi)$) representation of the displacement field in the basin satisfying the Helmholtz equation and the stress-free boundary condition (\ref{1-020}):
\begin{equation}\label{3-090}
u^{[1]}({\bf{x}})=\sum_{n=0}^{\infty}U_{n}^{[1]}(r)\cos(n\phi)=\sum_{n=0}^{\infty}A_{n}^{[1]}J_{n}(k^{[1]}r)\cos(n\phi)~;~\forall\mathbf{x}\in\Omega_{1}~.~
\end{equation}
I will encounter further on the functions $V_{n}^{[j]}(r)=U_{n,r}^{[j]}(r)$ and shall make use of the orthogonality relation
\begin{equation}\label{3-100}
\int_{\pi}^{2\pi}\cos(m\phi)\cos(n\phi)d\phi=\pi\delta_{mn}/\epsilon_{n}~;~m,n=0,1,2,....~,~
\end{equation}
whereby the $\phi$-integral of all products of functions of $r,\phi$ become  products of  of $U_{n}^{[j]}$ (or $U_{n}^{[j]*}$) with $U_{m}^{[j]}$ or $V_{m}^{[j]}$ respectively. Note that:
\begin{equation}\label{3-115}
V_{n}^{[0]}(r)=k^{[0]}\left(A_{n}^{[0]}\dot{J}_{n}(k^{[0]}r)+B_{n}^{[0]}\dot{H}_{n}(k^{[0]}r\right)~,~
\end{equation}
and
\begin{equation}\label{3-120}
V_{n}^{[1]}(r)=k^{[1]}A_{n}^{[0]}\dot{J}_{n}(k^{[1]}r)~,~
\end{equation}
wherein $\dot{J_{n}}(\zeta)=dJ_{n}/d\zeta$ and $\dot{H_{n}}(\zeta)=dH_{n}/d\zeta$.
\section{Employment of the SOV polar representations of the surface displacements and surface tractions in the conservation relation}
I make the following definitions (which employ the polar representations of the surface displacements and surface stresses):
\begin{multline}\label{4-010}
I^{[0]}_{\Gamma_{\infty}}=-\lim_{R\rightarrow\infty}\int_{\pi}^{2\pi} Rd\phi~u^{[0]*}(R,\phi)u_{,r}^{[0]}(R,\phi)~~,~~
I^{[0]}_{\Gamma_{b}}=\int_{\pi}^{2\pi} hd\phi~u^{[0]*}(h,\phi)u_{,r}^{[0]}(h,\phi)\\
I^{[1]}_{\Omega_{1}}=\Im\big[\big(k^{[1]}\big)^{2}\big]\int_{0}^{h}~rdr\int_{\pi}^{2\pi}~d\phi|u^{[1]}(r,\phi)|^{2}~~~~~~~~~~~~~~~~~~~~~~~~~~~~~~~~~~~~.
\end{multline}
whereby the conservation relation (\ref{2-145}) becomes
\begin{equation}\label{4-020}
\Re{\Bigg [}\frac{\mu^{[0]}}{\mu^{[1]}}{\Bigg ]}\Im I^{[0]}_{\Gamma_{\infty}}-\Im{\Bigg [}\frac{\mu^{[0]}}{\mu^{[1]}}{\Bigg ]}\Re I^{[0]}_{\Gamma_{b}}+I^{[1]}_{\Omega_{1}}=0~.
\end{equation}
The  orthogonality relations entail:
\begin{multline}\label{4-030}
I^{[0]}_{\Gamma_{\infty}}=\lim_{R\rightarrow\infty}\sum_{n=0}^{\infty}~\left(\frac{\pi R}{\epsilon_{n}}\right)
U_{n}^{[0]*}(R)V_{n}^{[0]}(R)~~,~~
I^{[0]}_{\Gamma_{b}}=\sum_{n=0}^{\infty}~\left(\frac{\pi h}{\epsilon_{n}}\right)U^{[0]*}(h)V_{n}^{[0]}(h)\\
I^{[1]}_{\Omega_{1}}=\Im\big[\big(k^{[1]}\big)^{2}\big]\sum _{n=0}^{\infty} \left(\frac{\pi}{\epsilon_{n}}\right)\int_{0}^{h}~rdr|U^{[1]}(r)|^{2}~~~~~.
\end{multline}
I take into account the facts that $J_{n}(k^{[0]}r)$ is real, $H_{n}(k^{[0]}r)=J_{n}(k^{[0]}r)+iY_{n}(k^{[0]}r)$ (wherein $Y_{n}(.)$ the Bessel function of the second kind) is complex, and $J_{n}(k^{[1]}r)$ is complex (because $k^{[1]}$ is complex) to enable (\ref{4-030}) to take the forms:
\begin{equation}\label{4-040}
I^{[0]}_{\Gamma_{\infty}}=-\lim_{R\rightarrow\infty}\sum_{n=0}^{\infty}~
\left(\frac{\pi k^{[0]}R}{\epsilon_{n}}\right)
\left(A_{n}^{[0]*}J_{n}(k^{[0]}R)+B_{n}^{[0]*}H_{n}^{*}(k^{[0]}R)\right)
\left(A_{n}^{[0]}\dot{J}_{n}(k^{[0]}R)+B_{n}^{[0]}\dot{H}_{n}(k^{[0]}R)\right)~,
\end{equation}
\begin{equation}\label{4-050}
I^{[0]}_{\Gamma_{b}}=\sum_{n=0}^{\infty}~
\left(\frac{\pi k^{[0]}h}{\epsilon_{n}}\right)
\left(A_{n}^{[0]*}J_{n}(k^{[0]}h)+B_{n}^{[0]*}H_{n}^{*}(k^{[0]}h)\right)
\left(A_{n}^{[0]}\dot{J}_{n}(k^{[0]}h)+B_{n}^{[0]}\dot{H}_{n}(k^{[0]}h)\right)~,
\end{equation}
\begin{equation}\label{4-060}
I^{[1]}_{\Omega_{1}}=\Im\big[\big(k^{[1]}\big)^{2}\big]|A_{n}^{[1]}|^{2}\left(\frac{\pi}{\epsilon_{n}}\right)
\int_{0}^{h}~J_{n}^{*}(k^{[1]}r)J_{n}(k^{[1]}r)rdr~.
\end{equation}
In order to deal with the integral
\begin{equation}\label{4-070}
\mathcal{I}_{n}=\int_{0}^{h}~
J_{n}^{*}(k^{[1]}r)J_{n}(k^{[1]}r)rdr~,
\end{equation}
I make use of formula (11.3.29) in \cite{as68} to obtain
\begin{equation}\label{4-080}
\mathcal{I}_{n}=\left(\frac{h}{\Im[(k^{[1]})^{2}]}\right)
\Im\left[k^{[1]}J_{n+1}(k^{[1]}h)J^{*}_{n}(k^{[1]}h)\right]~.
\end{equation}
so that
\begin{equation}\label{4-090}
I^{[1]}_{\Omega_{1}}=|A_{n}^{[1]}|^{2}\left(\frac{\pi}{\epsilon_{n}}\right)\Im\left[k^{[1]}hJ_{n+1}(k)J^{*}_{n}(k^{[1]}h)\right]
~.
\end{equation}
However, owing to formula (9.1.27) in \cite{as68}, it turns out that $\Im\left[\zeta J_{n}^{*}(\zeta)J_{n+1}(\zeta)\right]=-\Im\left[\zeta J_{n}^{*}(\zeta)\dot{J}_{n}(\zeta)\right]$ so that
\begin{equation}\label{4-095}
I^{[1]}_{\Omega_{1}}=-|A_{n}^{[1]}|^{2}\left(\frac{\pi}{\epsilon_{n}}\right)\Im\left[k^{[1]}h\dot{J}(k)J^{*}_{n}(k^{[1]}h)\right]
~.
\end{equation}

Furthermore, it is straightforward to show that $\Re I^{[0]}_{\Gamma_{b}}$ reduces to
\begin{equation}\label{4-100}
\Re I^{[0]}_{\Gamma_{b}}=-\sum_{n=0}^{\infty}~\left(\frac{\pi k^{[0]}h}{\epsilon_{n}}\right)
\left(a_{n}+b_{n}+c_{n}\right)~,
\end{equation}
wherein
\begin{equation}\label{4-110}
a_{n}=\left[|A_{n}^{[0]}\|^{2}+\|B_{n}^{[0]}|^{2}+2\Re\left(A_{n}^{[0]}B_{n}^{[0]*}\right)
\right]J_{n}(k^{[0]}h)\dot{J}_{n}(k^{[0]}h)~,
\end{equation}
\begin{equation}\label{4-120}
b_{n}=|B_{n}^{[0]}|^{2}Y_{n}(k^{[0]}h)\dot{Y}_{n}(k^{[0]}h)~,
\end{equation}
\begin{equation}\label{4-130}
c_{n}=-\Im\left[A_{n}^{[0]*}B_{n}^{[0]}J_{n}(k^{[0]}h)\dot{Y}_{n}(k^{[0]}h)-
A_{n}^{[0]}B_{n}^{[0]*}\dot{J}_{n}(k^{[0]}h)Y_{n}(k^{[0]}h)\right]~.
\end{equation}
with $Y_{n}(.)$ the Bessel function of the second kind.

Finally, I address the problem of the evaluation of $\Im I^{[0]}_{\Gamma_{\infty}}$. It is straightforward to reduce (\ref{4-040}) to
\begin{multline}\label{4-140}
I^{[0]}_{\Gamma_{\infty}}=
-\lim_{R\rightarrow\infty}\Im\sum_{n=0}^{\infty}~
\left(\frac{\pi k^{[0]}R}{\epsilon_{n}}\right)\times\\
\left[|B_{n}^{[0]}|^{2}H_{n}^{*}(k^{[0]}R)\dot{H}_{n}(k^{[0]}R)+
A_{n}^{[0]*}B_{n}^{[0]}J_{n}(k^{[0]}R)\dot{H}_{n}(k^{[0]}R)+B_{n}^{[0]*}A_{n}^{[0]}H_{n}^{*}(k^{[0]}R)\dot{J}_{n}(k^{[0]}R)\right]~.
\end{multline}
Further simplifications result from use of formula (9.1.17) in \cite{as68} which result in
\begin{multline}\label{4-150}
I^{[0]}_{\Gamma_{\infty}}=
-\lim_{R\rightarrow\infty}\Re\sum_{n=0}^{\infty}~
\left(\frac{1}{\epsilon_{n}}\right)\times\\
\left[
2|B_{n}^{[0]}|^{2}+\pi k^{[0]}R
\left(
A_{n}^{[0]*}B_{n}^{[0]}J_{n}(k^{[0]}R)\dot{Y}_{n}(k^{[0]}R)-B_{n}^{[0]*}A_{n}^{[0]}Y_{n}(k^{[0]}R)\dot{J}_{n}(k^{[0]}R)\right)\right]~,
\end{multline}
and the employment of the asymptotic forms in formulae (9.2.1) and (9.2.2) of \cite{as68} enables me to obtain
\begin{multline}\label{4-160}
I^{[0]}_{\Gamma_{\infty}}=
-\lim_{R\rightarrow\infty}\Re\sum_{n=0}^{\infty}~
\left(\frac{1}{\epsilon_{n}}\right)\times\\
\left[
2|B_{n}^{[0]}|^{2}+
\frac{\pi k^{[0]}R}{2}
\left(
A_{n}^{[0]*}B_{n}^{[0]}\left[\frac{4}{\pi k^{[0]}R}\right]\cos^{2}(\alpha_{n})+
B_{n}^{[0]*}A_{n}^{[0]}\left[\frac{4}{\pi k^{[0]}R}\right]\sin^{2}(\alpha_{n})
\right)
\right]~,
\end{multline}
wherein $\alpha_{n}=k^{[0]}R-n\pi/2-\pi/4$. The end result of taking the limit $R\rightarrow\infty$ is
\begin{equation}\label{4-170}
I^{[0]}_{\Gamma_{\infty}}=
-\sum_{n=0}^{\infty}~
\left(\frac{2}{\epsilon_{n}}\right)
\left[
|B_{n}^{[0]}|^{2}+\Re\left(A_{n}^{[0]}B_{n}^{[0]*}\right)
\right]~.
\end{equation}
Setting
\begin{equation}\label{4-180}
d=-\sum_{n=0}^{\infty}~\left(\frac{2}{\epsilon_{n}}\right)\Re\left(A_{n}^{[0]}B_{n}^{[0]*}\right)~,
\end{equation}
and dividing (\ref{4-170}) by $d$ results in
\begin{equation}\label{4-190}
\frac{I^{[0]}_{\Gamma_{\infty}}}{d}=1+
\sum_{n=0}^{\infty}~
\left(\frac{2}{\epsilon_{n}}\right)
|B_{n}^{[0]}|^{2}
~.
\end{equation}
so that the conservation relation takes the form
\begin{equation}\label{4-200}
-\Re\left[\frac{\mu^{[0]}}{\mu^{[1]}}\right]
\left[
[1+\sum_{n=0}^{\infty}~
\left(\frac{2}{\epsilon_{n}}\right)
|B_{n}^{[0]}|^{2}
\right]+
\Im\left[\frac{\mu^{[0]}}{\mu^{[1]}}\right]
\left(\frac{\Re I^{[0]}_{\Gamma_{b}}}{d}\right)-
\left(\frac{I^{[1]}_{\Omega_{1}}}{d}\right)=0~.
\end{equation}
However
\begin{equation}\label{4-210}
\Re\left[\frac{\mu^{[0]}}{\mu^{[1]}}\right]=
\frac{\mu^{[0]}\Re\mu^{[1]}}{|\mu^{[1]}|^{2}}~,~\Im\left[\frac{\mu^{[0]}}{\mu^{[1]}}\right]=
-\frac{\mu^{[0]}\Im\mu^{[1]}}{|\mu^{[1]}|^{2}}~\Rightarrow~
\frac{\Im\left[\frac{\mu^{[0]}}{\mu^{[1]}}\right]}{\Re\left[\frac{\mu^{[0]}}{\mu^{[1]}}\right]}=
~-\frac{\Im\mu^{[1]}}{\Re\mu^{[1]}}~,
\end{equation}
so that the final result ensues
\begin{equation}\label{4-220}
-\frac{\Im\mu^{[1]}}{\Re\mu^{[1]}}\left(\frac{\Re I^{[0]}_{\Gamma_{b}}}{d}\right)+
\frac{|\mu^{[1]}|^{2}}{\mu^{[0]}\Im\mu^{[1]}}\left(\frac{I^{[1]}_{\Omega_{1}}}{d}\right)+
\left(\frac{\sum_{n=0}^{\infty}\left(\frac{2}{\epsilon_{n}}\right)|B_{n}^{[0]}|^{2}}{d}\right)=1
~.
\end{equation}
This is the {\it explicit} form of the conservation law (for the semi-circular cylindrical basin seismic response problem) since three physics-related variable quantities (left side of the equation) must add up to the mathematical invariant quantity which, by the normalization procedure involving $d$, is unity (the right side of the equation). This is not a conservation of energy relation because  the entries of the left or right side of the equations were established in a
space-frequency (rather than space-time) framework and therefore are not (normalized) energies. Nevertheless, the entries on the left-hand side of the equation do
have a physical origin (somehow related to energy flow) so that I  call them "normalized output fluxes". Similarly, I call the right hand side the "normalized input flux". More specifically: I call the term related to  $I^{[0]}_{\Gamma_{b}}$ the "normalized interface flux" since it involves only the displacement and stress at the interface $\Gamma_{b}$, I call  the term related to $\Omega_{1}$ the "normalized absorption flux" since it clearly has to do with the volumic absorption within the basin, and I call the term related to $\sum |B_{n}^{[0]}|^{2}$ the "normalized scattering flux" since the $B_{n}^{[0]}$ are the amplitudes of the scattered field (in the energy context, this would be related to what is frequently called "the radiation damping term" \cite{mw94}).

Alternatively, the problem at hand appears to be related to that of a black box submitted to some sort of solicitation and what goes into the box must equal what is absorbed therein, plus what stays at the surface of the box,  plus what comes out of the box. I call what comes in the "gain" ($G$), what is absorbed the "volume loss" ($L_{v}$), what stays at the surface ($L_{s}$) the "surface loss", and what comes out the "radiation loss" ($L_{r}$). Clearly, the sum of these three losses must equal the gain. which is what the conservation relation states, since the latter is expressed by
\begin{equation}\label{4-230}
L_{s}+L_{v}+L_{r}=G
~,
\end{equation}
wherein
\begin{equation}\label{4-240}
L_{s}=-\frac{\Im\mu^{[1]}}{\Re\mu^{[1]}}\left(\frac{\Re I^{[0]}_{\Gamma_{b}}}{d}\right)~,~
L_{v}=\frac{|\mu^{[1]}|^{2}}{\mu^{[0]}\Im\mu^{[1]}}\left(\frac{I^{[1]}_{\Omega_{1}}}{d}\right)~,~
L_{r}=\left(\frac{\sum_{n=0}^{\infty}\left(\frac{2}{\epsilon_{n}}\right)|B_{n}^{[0]}|^{2}}{d}\right)~,~
G=1~~.
\end{equation}

Recall that this conservation relation was established by employing the sole field {\it representations}. The next steps are: 1) to show  what the different loss terms look like when the representations are replaced by the actual field {\it solutions} and 2) to show that these solutions are such that the conservation relation is satisfied. Step 2) is the means by which one can verify that the solutions are correct (or, in any case, not inconsistent, since the real test of quality of a solution is that it is such that the boundary and continuity conditions are satisfied).
\section{Exact resolution of the problem by employment of the SOV polar representations of the surface displacements and surface tractions in the continuity relations}
The polar coordinate forms of the continuity of surface displacement and surface traction relations (\ref{bc-040})-(\ref{bc-050})
are:
\begin{equation}\label{5-010}
u^{[0]}(h,\phi)=u^{[1]}(h,\phi)~;~\forall\phi\in[\pi,2\pi]~,~
\end{equation}
\begin{equation}\label{5-020}
\mu^{[0]}u^{[0]}_{,r}(h,\phi)=\mu^{[1]}u_{,r}^{[1]}(h,\phi)~;~\forall\phi\in[\pi,2\pi]~.~
\end{equation}
The introduction therein of (\ref{3-090})-(\ref{3-120}) gives rise to the system of two linear equations
\begin{equation}\label{5-030}
U_{n}^{[0]}(h)-U_{n}^{[1]}(h)=0~;~n=0,1,2,...~,~
\end{equation}
\begin{equation}\label{5-040}
\mu^{[0]}V_{n}^{[0]}(h)-\mu^{[1]}V_{n}^{[1]}(h)=0~;~n=0,1,2,...~,~
\end{equation}
the solution of which is
\begin{equation}\label{5-050}
B_{n}^{[0]}=A_{n}^{[0]}\left[\gamma^{[1]}J_{n}(k^{[0]}h)\dot{J}_{n}(k^{[1]}h)-
\gamma^{[0]}J_{n}(k^{[1]}h)\dot{J}_{n}(k^{[0]}h)\right]/D_{n}~;~n=0,1,2,...~,~
\end{equation}
\begin{multline}\label{5-060}
A_{n}^{[1]}=A_{n}^{[0]}\left[-\gamma^{[0]}\dot{J}_{n}(k^{[0]}h)H_{n}(k^{[0]}h)+
\gamma^{[0]}J_{n}(k^{[0]}h)\dot{H}_{n}(k^{[0]}h)\right]/D_{n}=\\
A_{n}^{[0]}\left[\frac{2i\gamma^{[0]}}{\pi k^{[0]}h D_{n}}\right]~;~n=0,1,2,...~,~
\end{multline}
wherein $\gamma^{[j]}=k^{[j]}\mu^{[j]}~;~j=0,1$ and
\begin{equation}\label{5-070}
D_{n}=\gamma^{[0]}\dot{H}_{n}(k^{[0]}h)J_{n}(k^{[1]}h)-
\gamma^{[1]}H_{n}(k^{[0]}h)\dot{J}_{n}(k^{[1]}h)
~;~n=0,1,2,...~.
\end{equation}
is the determinant of the two-by-two linear system (\ref{5-030})-(\ref{5-040}) for each $n$.

In the next section I shall demonstrate that these solutions  numerically lead to the satisfaction of the conservation (of flux or gain/loss)  relation (\ref{4-230}) {\it for all $n\ge 0$} and are therefore consistent (in fact, it is easily verified that they are also correct in that they, together with the field representations,  are such as to satisfy the  partial differential equation plus the boundary and continuity conditions of the problem).
\section{Computational evidence of resonances}\label{numerics}
In all the figures which follow, the numerical computations on which they are based, apply to the parameter choices: $\beta^{[0]}=1000ms^{-1}$,  $\rho^{[0]}=2200kgm^{-3}$, $\beta_{ref}^{[1]}=200ms^{-1}$, $Q^{[1]}=25$, $\rho^{[1]}=1800kgm^{-3}$, $h=100m$ and $A^{i}=1$ for all frequencies $f$. $f$ and $\theta^{i}$ will be variables.

\begin{figure}[ht]
  \begin{center}aa1
    \subfloat[]{
      \includegraphics[width=0.45\textwidth]{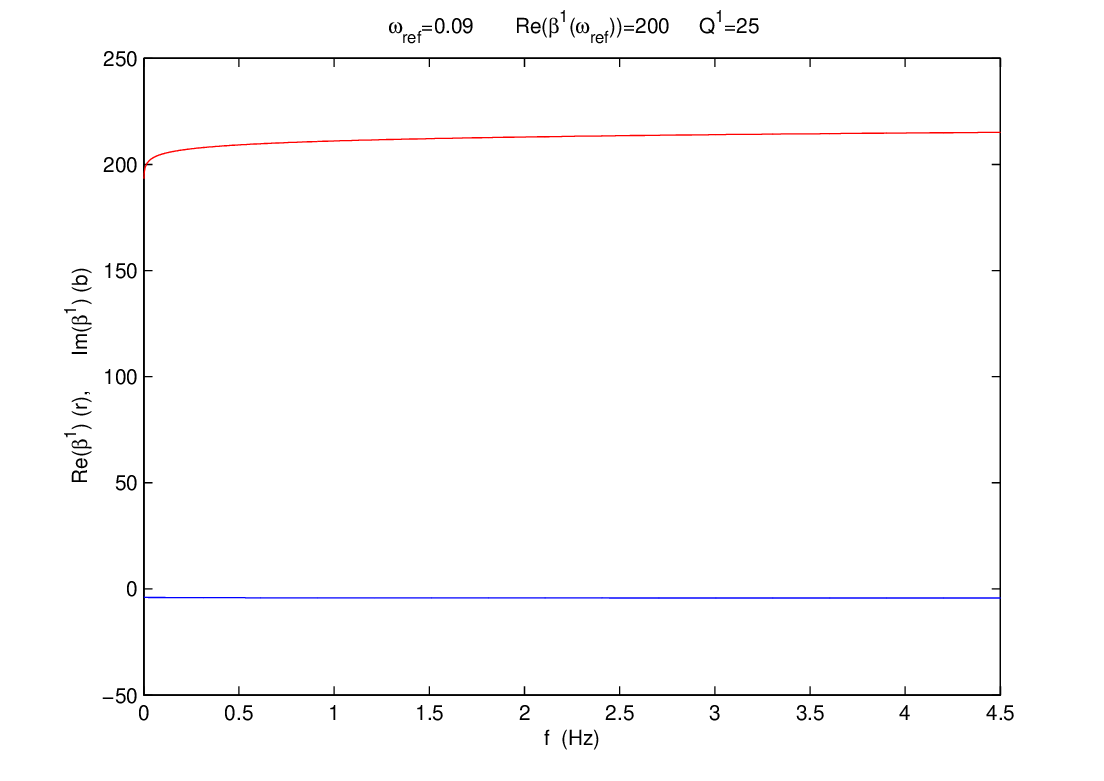}
      \label{a}
                         }
    \subfloat[]{
      \includegraphics[width=0.45\textwidth]{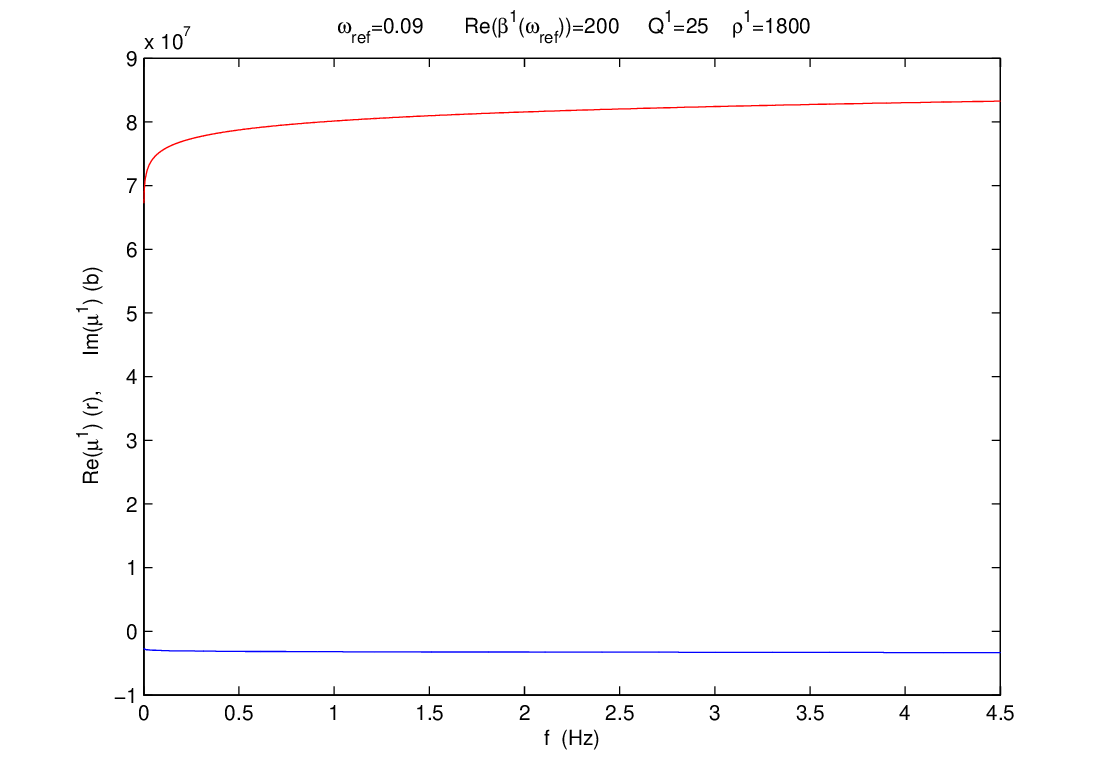}
      \label{b}
                         }
    \caption{
(a) $\Re(\beta^{[1]})$ vs $f$ (red), $\Im(\beta^{[1]})$ vs $f$ (blue).\\
(b) $\Re(\mu^{[1]})$ vs $f$ (red), $\Im(\mu^{[1]})$ vs $f$ (blue).
}
    \label{dispersion}
  \end{center}
\end{figure}

The dispersive nature of the sediment medium is accounted-for by (\ref{1-002}) whose numerical translation (using the above-mentioned parameters) is given in the fig.\ref{dispersion} below

All the quantities of interest (i.e., $u^{[0]}$, $u^{[1]}$, $L_{s}$, $L_{v}$, $L_{r}$) contain series of the form $\sum_{n=0}^{\infty}$ which, for numerical purposes, are necessarily truncated to $\sum_{n=0}^{Nmax}$. Thus, the first task is to determine how the choice of $Nmax$ influences the numerical values of the quantities of interest. Since the two displacement functions largely-condition the three loss functions, the convergence of the latter with respect to $Nmax$ must reflect the way  the displacement functions converge with respect to $Nmax$. Thus, I examine how the sole loss functions vary with $Nmax$, assuming that the upper limit of the series for $u^{[0]}$ is the same as that for $u^{[1]}$.
\subsection{Variation of $Nmax$ in the three loss functions and verification of the conservation relation}
The following two figures contain three panels, each of which depicts (in ordinates): $L_{s}$ (red), $L_{v}$ (blue), $L_{r}$ (magenta) and $L_{s}+L_{v}+L_{r}$ (green) as a function of frequency $f$ in Hz (abscissas). The sum  $L_{s}+L_{v}+L_{r}$ should equal $G=1$ if energy (flux) is conserved. The first figure applies to $Nmax=0,1,2$ and the second figure to $Nmax=2,3,4$ and in both figures the angle of incidence is $\theta^{i}=60^{\circ}$.

\newpage
\begin{figure}[ht]
  \begin{center}
    \subfloat[]{
      \includegraphics[width=0.3\textwidth]{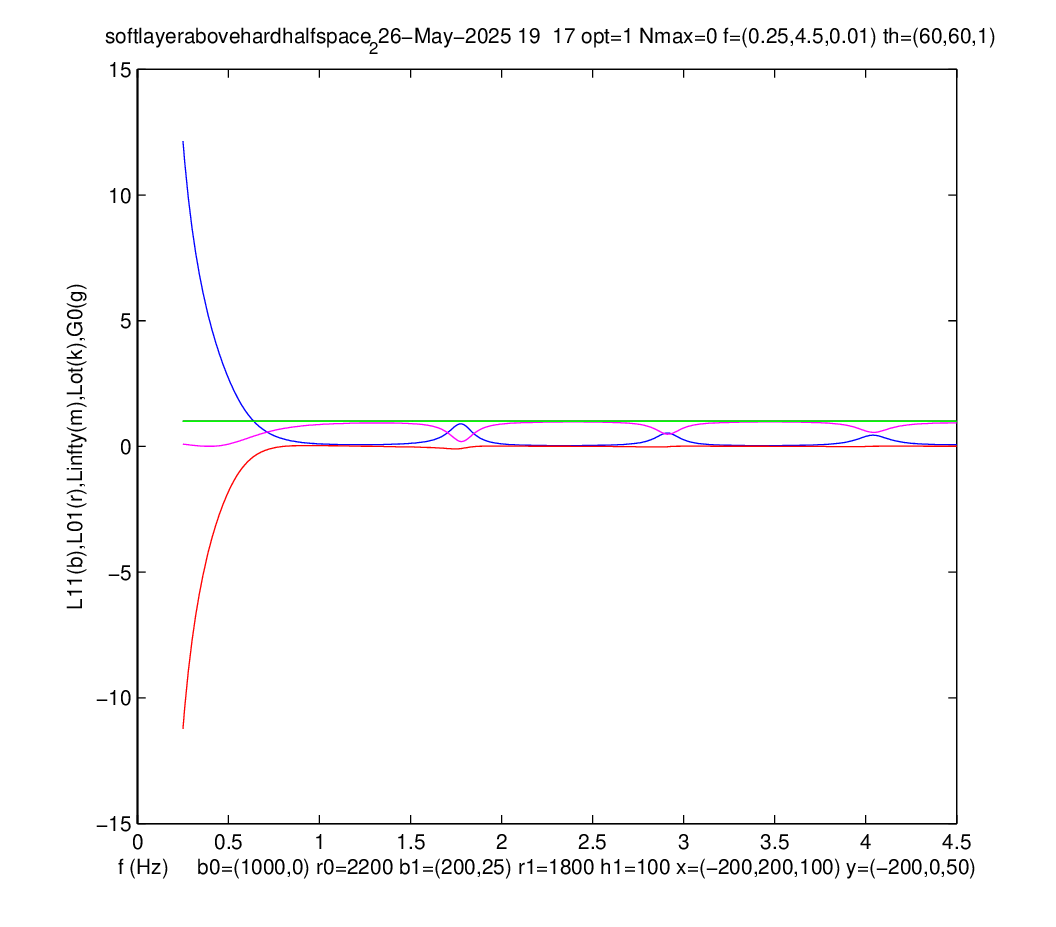}
      \label{a}
                         }
    \subfloat[]{
      \includegraphics[width=0.3\textwidth]{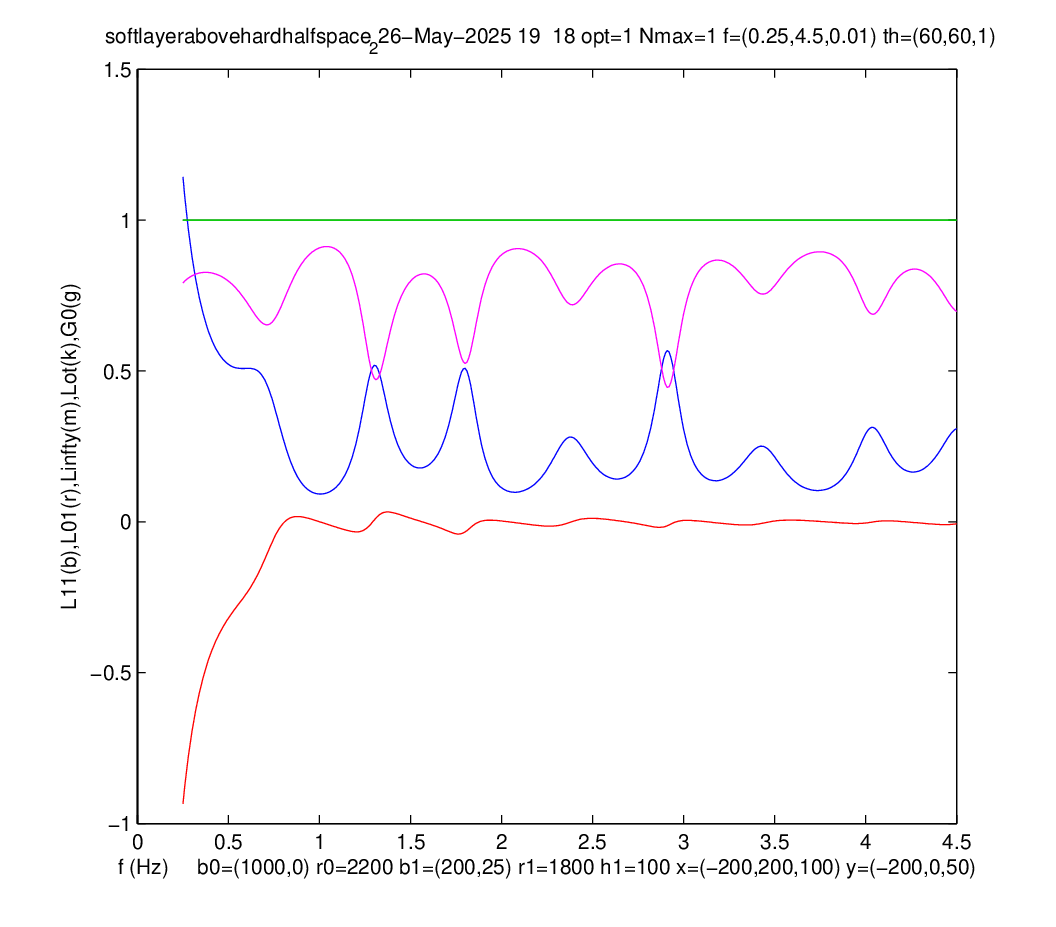}
      \label{b}
                         }
    \subfloat[]{
      \includegraphics[width=0.3\textwidth]{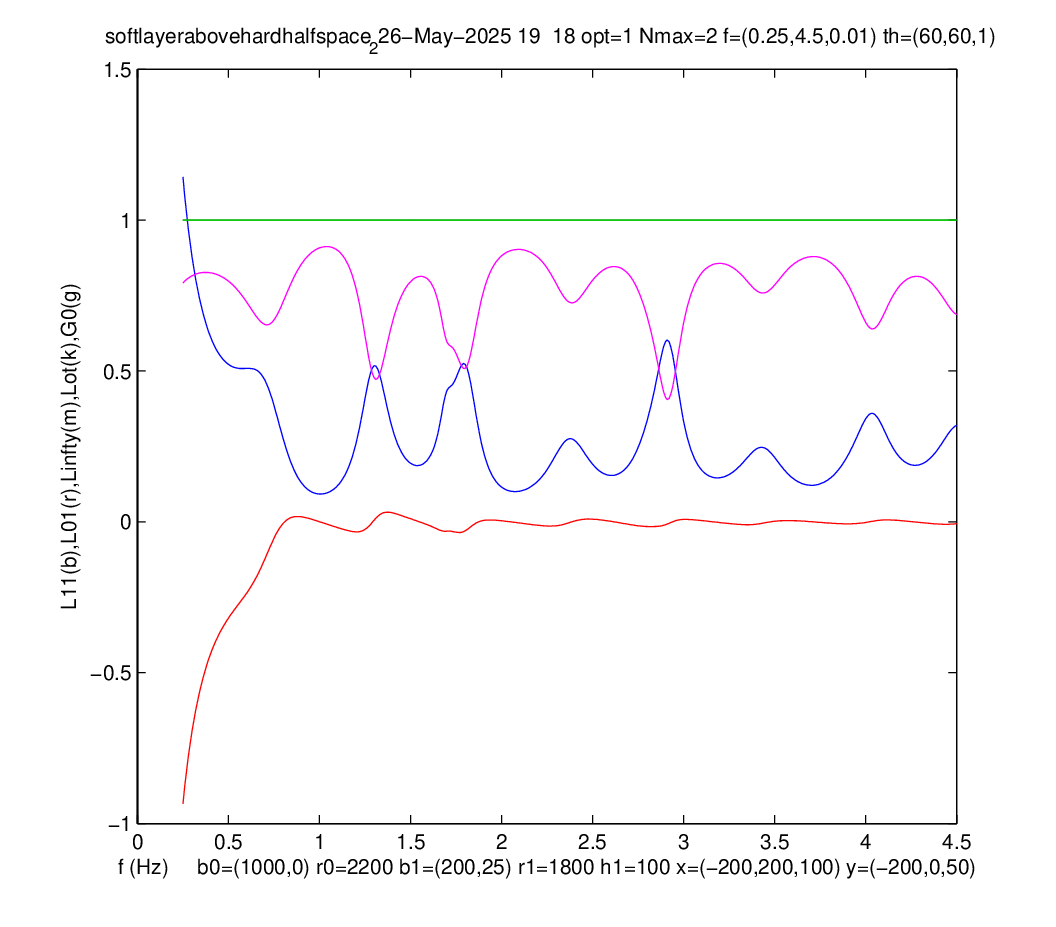}
      \label{b}
                         }

    \caption{Loss functions and their sum vs $f$(Hz):
(a) $Nmax=0$, (b) $Nmax=1$, (c) $Nmax=2$.
}
    \label{vary Nmax 0-2}
  \end{center}
\end{figure}
\begin{figure}[ht]
  \begin{center}
    \subfloat[]{
      \includegraphics[width=0.3\textwidth]{softbasinwithinhardhalfspace_2-4.eps}
      \label{a}
                         }
    \subfloat[]{
      \includegraphics[width=0.3\textwidth]{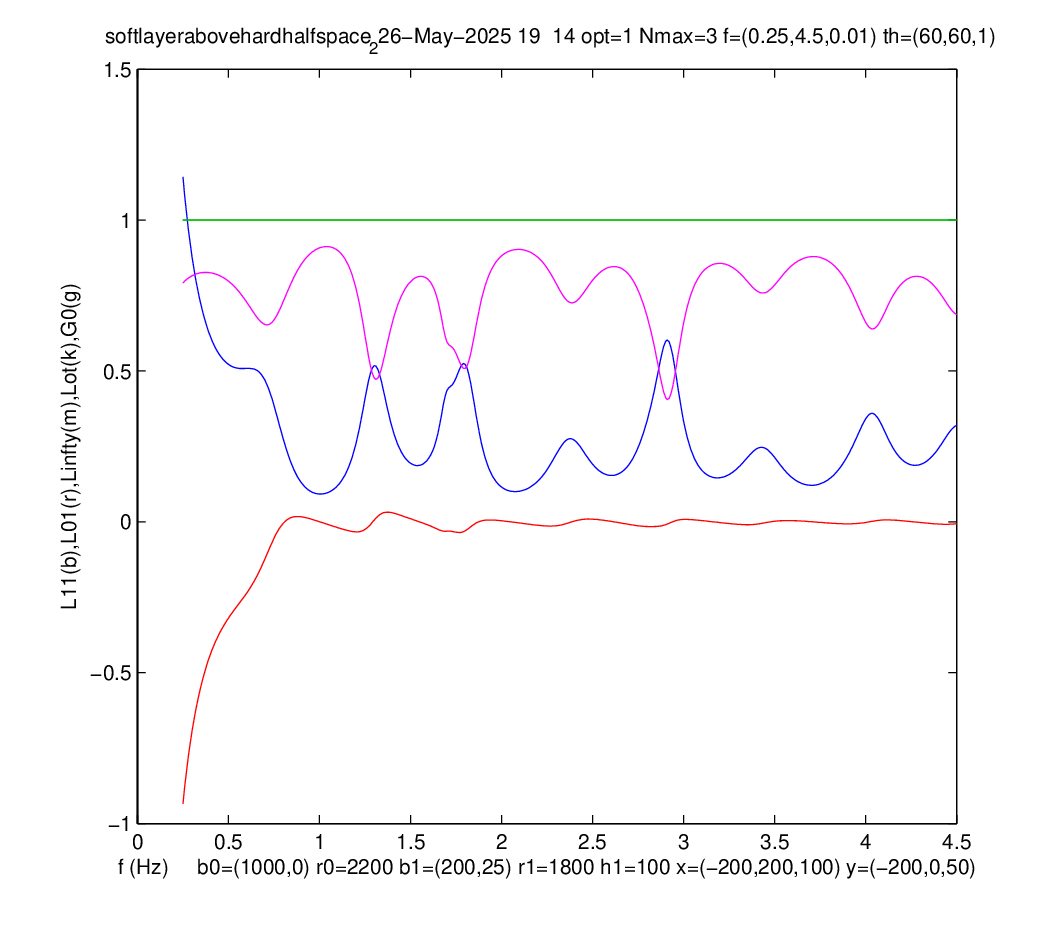}
      \label{b}
                         }
    \subfloat[]{
      \includegraphics[width=0.3\textwidth]{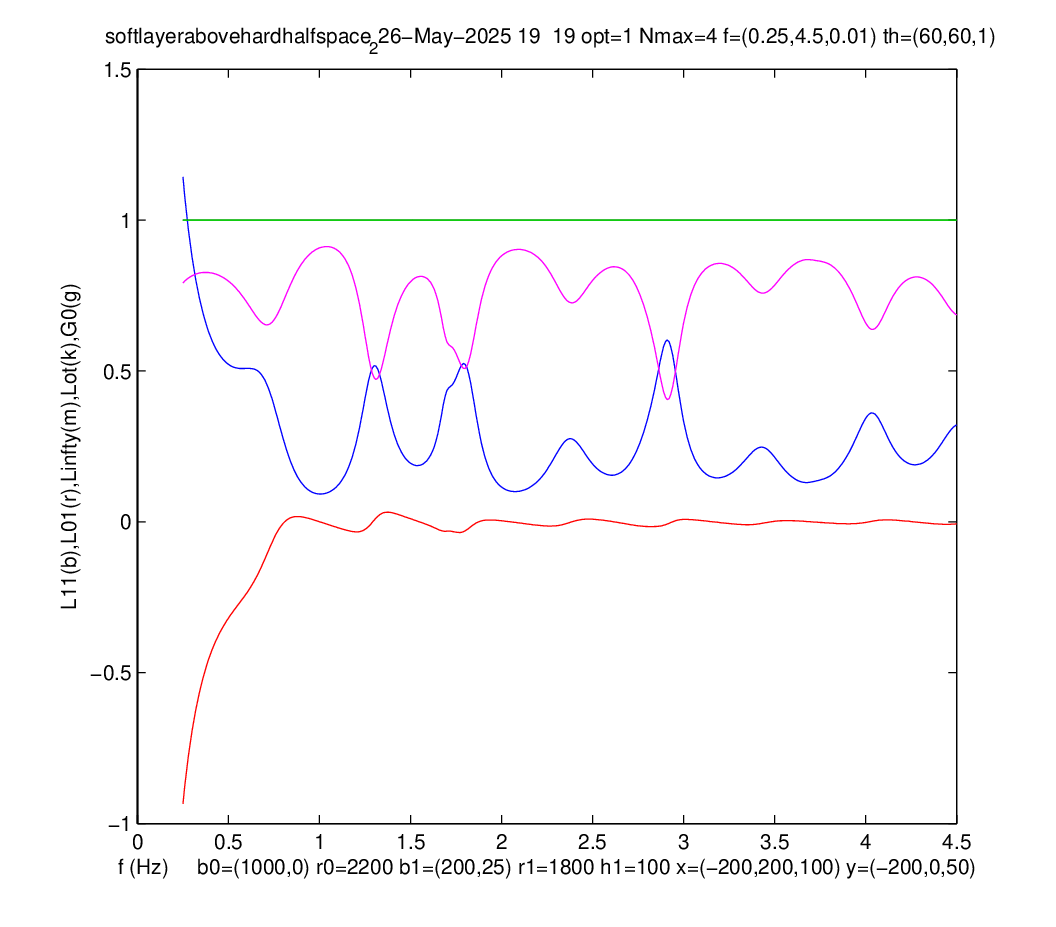}
      \label{b}
                         }

    \caption{Loss functions and their sum vs $f$(Hz):
(a) $Nmax=2$, (b) $Nmax=3$, (c) $Nmax=4$.
}
    \label{vary Nmax 2-4}
  \end{center}
\end{figure}

These two figures suggest that:\\
1) the behavior of the the three loss function is anomalous at very low $f$, but their sum is not anomalous at these frequencies\\
2) energy (flux) is conserved at all $f$ and for all $Nmax$\\
3) convergence is rapid with respect to increasing $Nmax$\\
4) $L_{r}$ is generally greater than $L_{v}$ except in the neighborhood of what appear to be  resonance  frequencies, while $L_{s}$ is small and hardly-varying with $f$ except for very small $f$.
\clearpage
\newpage
\subsection{Frequency of occurrence of resonances}
The material I am about to expose here is offered to deal with the first and fourth remarks of the previous section and appeals largely to what I wrote in my previous publication \cite{wi95}.

From its definition, I must conclude that $L_{v}$ is large when the displacement $u$ is large at the majority of locations within the basin, and I showed in the previous section that this occurs only at certain frequencies, which I called 'resonance frequencies'. A necessary, but not sufficient, condition for this to occur at a frequency $f_{r}$ is that the $n$-th coefficient $A_{n}^{[1]}$ be large at $f_{r}$, and, for this to occur, (almost) regardless of the other characteristics of the seismic load, the  modulus of the determinant $|D_{n}|$ must be nearly or totally equal to zero at $f_{r}$.

In \cite{wi95} I dealt {\it analytically} with the issue of finding the zeros of $|D_{n}|$; here I find it easier to do this by looking {\it numerically} for the maxima of $1/|D_{n}|$.

\begin{figure}[ht]
\begin{center}
\includegraphics[width=0.8\textwidth]{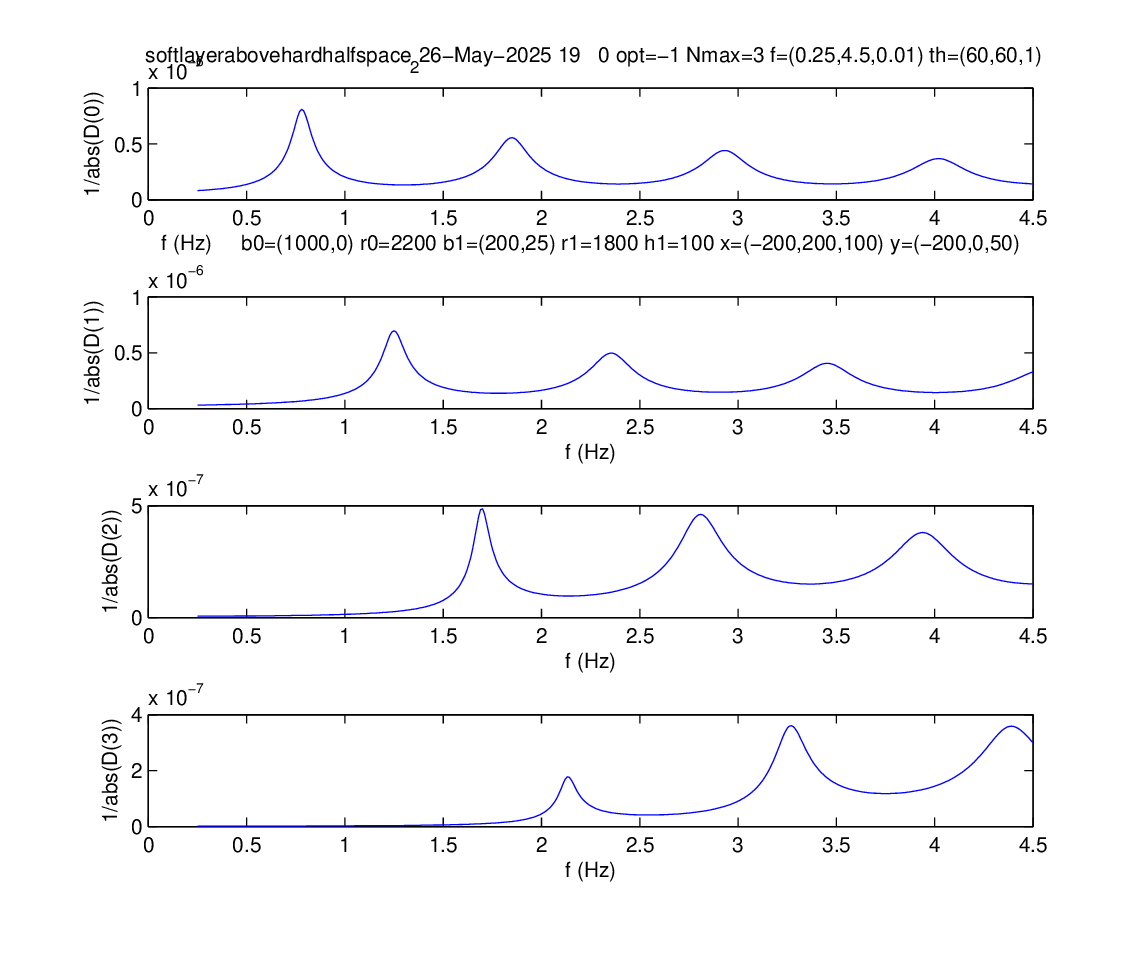}
\caption{ $1/|D_{n}|$(a.u.) vs $f$(Hz) for $n=0,1,2,3$}
\label{softcircbasin6}
\end{center}
\end{figure}
The result of such a procedure is given in fig.\ref{softcircbasin6} wherein $1/|D_{n}|$ is plotted against $f$ for $n=0,1,2,3$ (going from the top to the bottom of the figure). The hypothesis of the successive peaks being the sign of the existence of resonances is fortified by the fact that increasing (as is done in \cite{wi95}) $Q^{[1]}$ sharpens the peaks without modifying substantially their location. Consequently, the plausible interpretation of the $l-$th subfigure of fig.\ref{softcircbasin6} is that the position of the first peak therein corresponds to that of the fundamental resonance frequency $f_{l0}$, that of the second peak to the first overtone resonance frequency $f_{l1}$, that of the third peak to the second overtone resonance frequency $f_{l2}$,......, of the $l-$th partial wave. The numerical values of the eigenfrequencies, taken from  fig.\ref{softcircbasin6}, are (in Hz),:\\
$f_{00}$=0.776,~~$f_{01}$=1.84,~~$f_{02}$=2.93,~~$f_{04}$=4.01,\\
$f_{10}$=1.25 ,~~$f_{11}$=2.35,~~$f_{12}$=3.47,\\
$f_{20}$=1.70 ,~~$f_{21}$=2.81,~~$f_{22}$=3.95,\\
$f_{30}$=2.13 ,~~$f_{31}$=3.27,~~$f_{32}$=4.39.

\begin{figure}[ht]
  \begin{center}
    \subfloat[]{
      \includegraphics[width=0.45\textwidth]{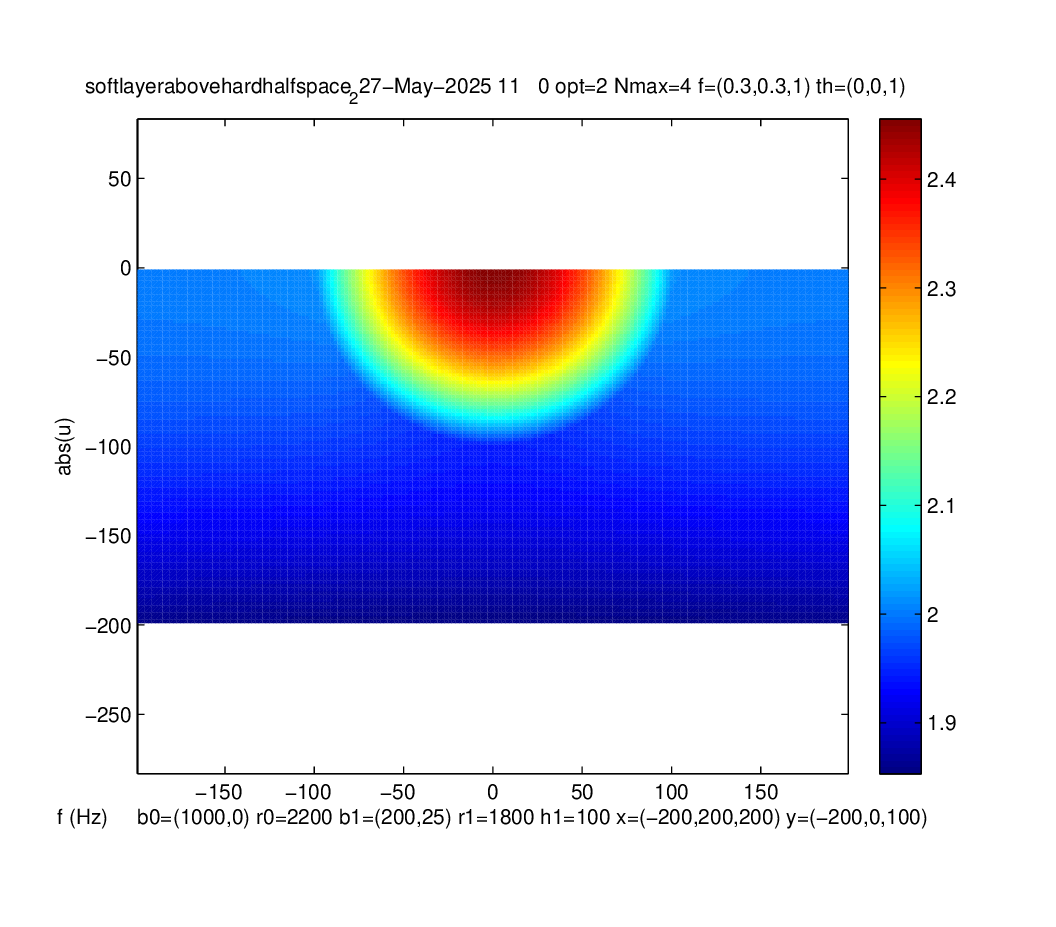}
      \label{a}
                         }
    \subfloat[]{
      \includegraphics[width=0.45\textwidth]{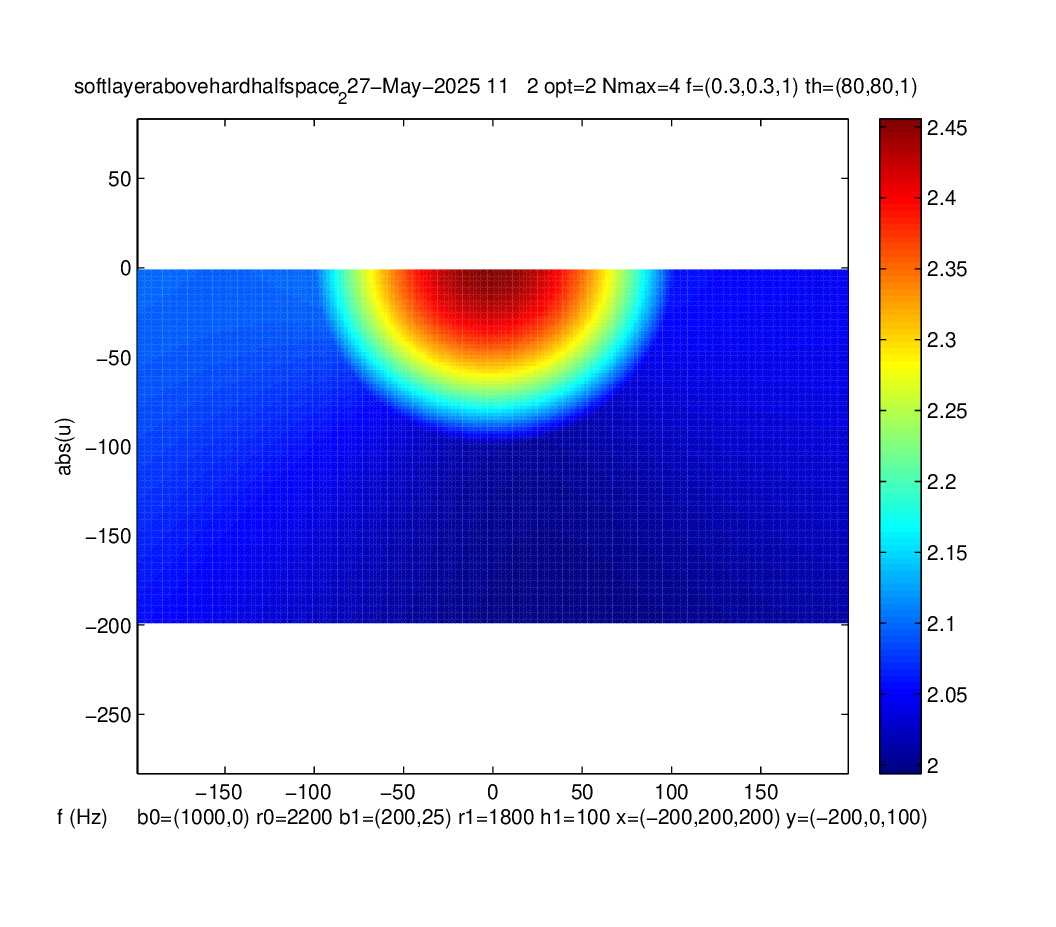}
      \label{b}
                         }
    \caption{Map of $|u^{j}(x,y)|; j=0,1$ for two seismic loads at $f=0.3$Hz:
(a) $\theta^i=0^{\circ}$.  (b) $\theta^i=80^{\circ}$.
}
    \label{softcircbasin7-8}
  \end{center}
\end{figure}

An interesting feature of fig.\ref{softcircbasin6} is that no resonances occur for very low frequencies (i.e., $f$ approximately smaller than 0.75Hz in this example), which means that the anomalous behavior of the loss functions, alluded-to in the first remark of the previous section, is not due to a resonance. This is substantiated by the plots of the modulus of the displacement field plots in fig.\ref{softcircbasin7-8} (for $\theta^{i}=0^{\circ}$ on the left and $\theta^{i}=80^{\circ}$ on the right) which exhibit a uniformly-weak response (not only within, but also outside, the basin) and otherwise nothing unusual at this non-resonant very low frequency $f=0.3$Hz. This shows that the anomalous behavior of the loss functions at very low frequencies is not due to an error in the computation of the displacement. Otherwise, I have no explanation of, nor am able to offer a solution for, this anomalous behavior other than to ignore it (for very low frequencies).

The more reassuring feature of the last of figs.\ref{vary Nmax 2-4} compared to fig.\ref{softcircbasin6} is that the positions  of the volumic absorption peaks of the former correlate quite well with the positions of at least one of the resonance peaks of the latter, which fact substantiates my hypothesis that the volumic aborption peaks are due to resonances in the basin response.
\section{Relation of resonant response to the excitation of an eigenmode}
%
\subsection{Mathematical definition and properties of an eigenmode}
To give more credit to the last statement above, I now return to the mathematical aspects of the problem (relying, as before, on \cite{wi95}). Strictly speaking, a $l-$th order partial wave resonance occurs at frequencies $f_{lm}~;~m=0,1,2,...$ ($m=0$ corresponds to the fundamental resonance frequency and $m>0$ to the overtones) for which
\begin{equation}\label{8-010}
D_{l}=\gamma^{[0]}\dot{H}_{l}(k^{[0]}h)J_{l}(k^{[1]}h)-
\gamma^{[1]}H_{l}(k^{[0]}h)\dot{J}_{l}(k^{[1]}h)=0~,
\end{equation}
whence
\begin{equation}\label{8-020}
\gamma^{[1]}\dot{J}_{l}(k_{lm}^{[1]}h)=
\gamma^{[0]}\frac{\dot{H}_{l}(k_{lm}^{[0]}h)}{H_{l}(k_{lm}^{[0]}h)}J_{l}(k_{lm}^{[1]}h)~,
\end{equation}
wherein $k_{lm}^{[j]}=2\pi f_{lm}/\beta^{[j]}$. It follows that
\begin{equation}\label{8-030}
B_{l}^{[0]}=A_{l}^{[1]}\frac{J_{l}(k_{lm}^{[1]}h)}{H_{l}(k_{lm}^{[0]}h)}~,
\end{equation}
at a $l-$th order partial wave resonance (eigen-) frequency. When this occurs, the contribution of the other partial waves in the series expressions of the displacements is negligible, which fact translates to
\begin{equation}\label{8-040}
u^{s}(\mathbf{x})=B_{l}^{[0]}u_{lm}^{+}(\mathbf{x})=
A_{l}^{[1]}\frac{J_{l}(k_{lm}^{[1]}h)}{H_{l}(k_{lm}^{[0]}h)}H_{l}(k_{lm}^{[0]}r)\cos(l\phi)~,
\end{equation}
\begin{equation}\label{8-050}
u^{[1]}(\mathbf{x})=A_{l}^{[1]}u_{lm}^{-}(\mathbf{x})=
A_{l}^{[1]}J_{l}(k_{lm}^{[1]}r)\cos(l\phi)~,
\end{equation}
at a $lm-$th order partial wave resonance (eigen-) frequency.

The couple $u^{+}_{lm}(r,\phi)=\frac{J_{l}(k_{lm}^{[1]}r)}{H_{l}(k_{lm}^{[0]}h)}
H_{l}(k_{lm}^{[0]}r)\cos(l\phi)$,
 $u^{-}_{lm}(r,\phi)=J_{l}(k_{lm}^{[1]}r)\cos(l\phi)$ constitutes what may be called the $lm-$th order eigenmode of the (soft basin within a hard halfspace) configuration since the induced motion (i.e., $u^{s}$ in the bedrock and $u^{[1]}$ within the basin) in the latter reduces to that of the the $lm-$th order eigenmode when the frequency $f$ coincides with the eigenfrequency $f_{lm}$. It
is important to realize that this is only true when $f=f_{lm}$. Moreover, the eigenmode has the peculiar property of satisfying the continuity of stress and displacement conditions on the lower boundary of the basin (although only at $f=f_{lm}$). This results from the following demonstration:
\begin{equation}\label{8-060}
u_{lm}^{+}(h,\phi)-u_{lm}^{-}(h,\phi)=\left(\frac{J_{l}(k_{lm}^{[1]}r)}{H_{l}(k_{lm}^{[0]}h)}\right)
H_{l}(k_{lm}^{[0]}h)\cos(l\phi)-J_{l}(k_{lm}^{[1]}r)\cos(l\phi)=0~;~\forall \phi\in[\pi,2\pi]~,
\end{equation}
\begin{multline}\label{8-070}
\mu^{[0]}u_{lm,r}^{+}(h,\phi)-\mu^{[1]}u_{lm,r}^{-}(h,\phi)=
\gamma^{[0]}\left(\frac{J_{l}(k_{lm}^{[1]}h)}{H_{l}(k_{lm}^{[0]}h)}\right)
\dot{H}_{l}(k_{lm}^{[0]}h)\cos(l\phi)-\gamma^{[1]}\dot{J}_{l}(k_{lm}^{[1]}h)\cos(l\phi)=\\
\frac{D_{l}}{H_{l}(k_{lm}^{[0]}h)}\cos(l\phi)=
\frac{0}{H_{l}(k_{lm}^{[0]}h)}\cos(l\phi)=0~;~\forall\phi\in[\pi,2\pi]~.
\end{multline}

All this means that it can be expected that the map, at $f=f_{lm}$, of the displacement field within the basin will essentially be the map of the the $u_{lml}^{-}$ component of $lm-$th eigenfunction. This will be verified further on where it will be shown that such an eigenfunction is characterized by a spatially-alternating series of maxima and minima, the number of which increases with increasing eigenfrequency.
\clearpage
\newpage
\subsection{How the incident angle affects resonant response}
Until now I eluded the question of how the incident angle $\theta^{i}$ affects resonant response, although I did stress (also in \cite{wi95}) the fact that the eigenfrequencies do not depend on $\theta^{i}$ (nor, for that matter, on $A^{i}$). What I did not mention was that this stance is not unanimously adopted in the seismological community since, for instance, in the conclusion of \cite{wd95} appear statements such as: "The steady state results show that the incident angle changes the amplitude and fundamental frequency for all P-, SV- and SH-waves. The fundamental frequency is also altered as the incident angle changes". Such statements merit further computational and theoretical attention which I provide hereafter.

As is commonplace, the authors of \cite{wd95} base their statements on comparisons of seismic displacement on the ground for different $\theta^{i}$. This procedure raises the  question:  is there  a most suitable location on the ground at which this comparison be made or must the comparison be made at more than one (how many?) ground locations? A second question is: what conclusions can be made if it is found (see the graphs in the next section of basin displacement (which include ground displacement)) that ground response, embodied in the modulus of the ground displacement, can apparently attain any value from zero to infinity? For this reason,  I think it to be preferable to do the comparison via the sole volumic loss $I_{v}$ function whose values are known a priori to be situated between 0 and 1.

\begin{figure}[ht]
  \begin{center}
    \subfloat[]{
      \includegraphics[width=0.3\textwidth]{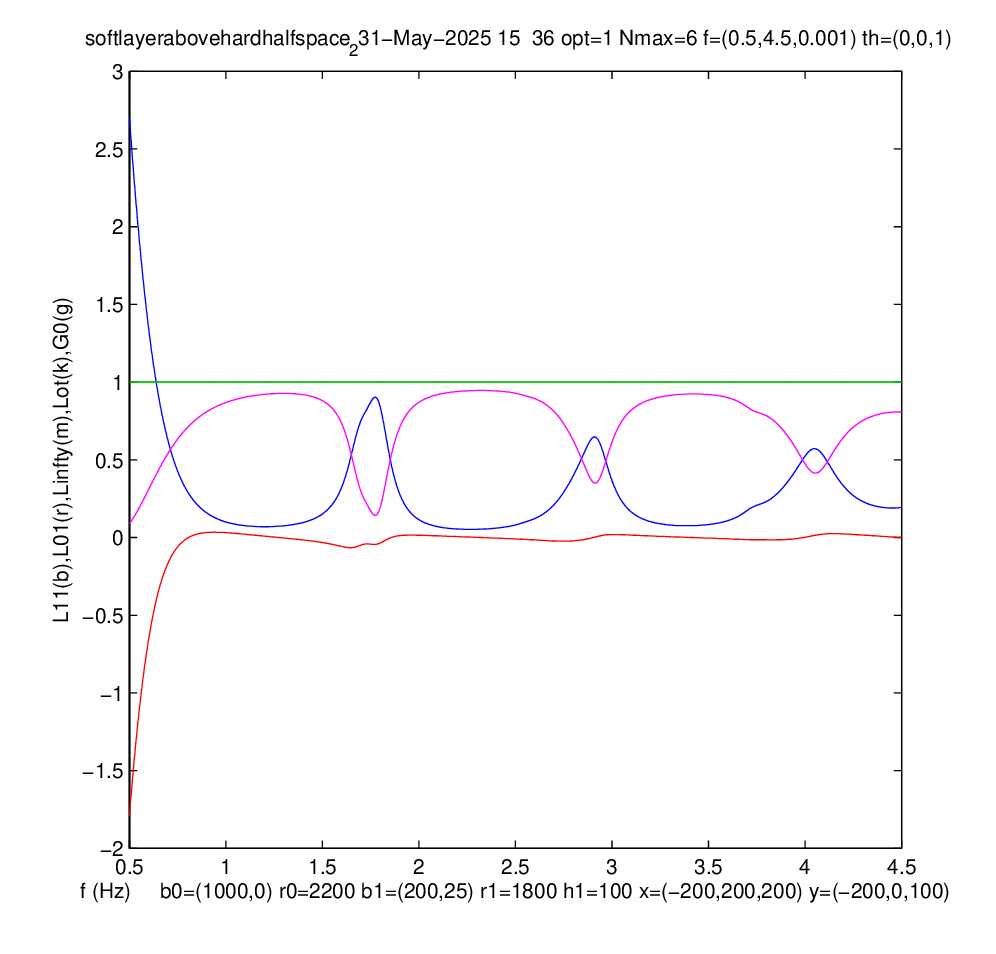}
      \label{a}
                         }
    \subfloat[]{
      \includegraphics[width=0.3\textwidth]{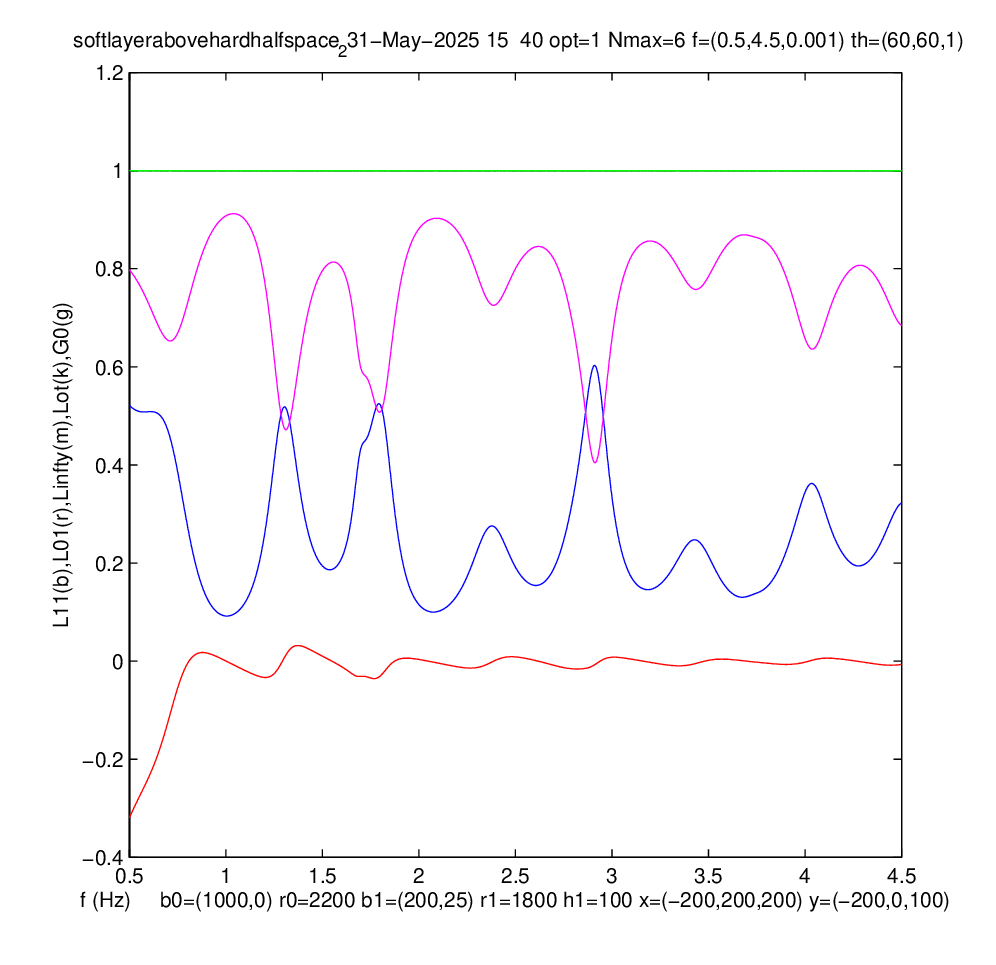}
      \label{b}
                         }
    \subfloat[]{
      \includegraphics[width=0.3\textwidth]{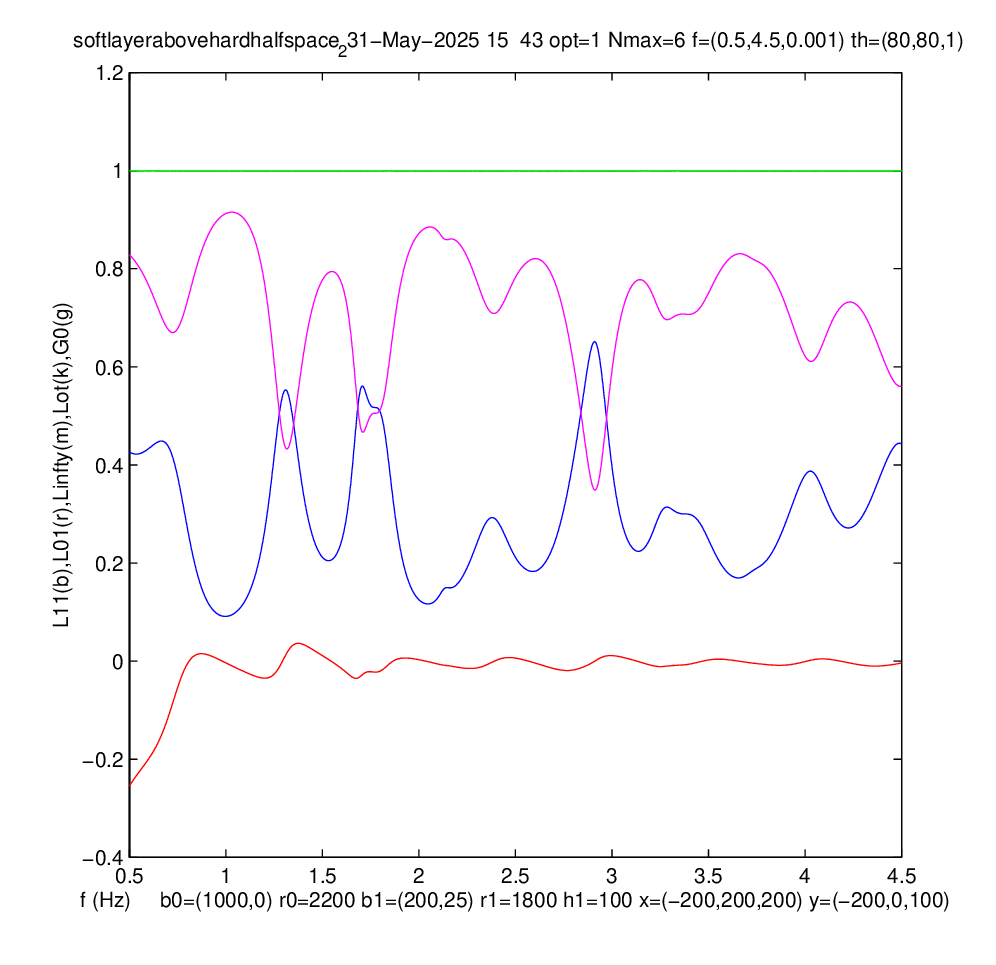}
      \label{b}
                         }

    \caption{Loss functions and their sum vs $f$(Hz) for different seismic loads:\\
(a) $\theta^i=0^{\circ}$,  (b) $\theta^i=60^{\circ}$,  (c) $\theta^i=80^{\circ}$.
}
    \label{vary thetai}
  \end{center}
\end{figure}

The  result of such a comparison is given in fig.\ref{vary thetai}. All the constitutive parameters are as previously and the choice $Nmax=6$ was made to ensure convergence of all the series involved in the computations. The three panels below are relative to $\theta^{i}=0^{\circ}$, $60^{\circ}$ and $80^{\circ}$.
As previously, the blue curves are relative to $I_{v}$, and it is on them that I focus my attention. The  attentive reader will remark that: 1) there is a strong resemblance between the low frequency (this includes the fundamental resonance frequency $f_{00}=0.776$Hz of the $lm=00$ eigenfunction) behavior of the blue curve in panel (a) of fig.\ref{vary thetai} and its counterpart in fig.\ref{vary Nmax 0-2} although the former applies to $\theta^{i}=0^{\circ}$ (and $Nmax=6$) and the latter to $\theta^{i}=60^{\circ}$ (and $Nmax=0$) and 2) this resemblance diminishes as the frequency increases. A plausible explanation of these remarks is that (only) the low-frequency response is dominated by the $n=0$ partial wave (in the partial wave series representation of the displacement field in the basin) and that this low-frequency response is insensitive to  the choice of $\theta^{i}$. This is better understood if I recollect what these (and particularly the one relative to the basin domain) partial wave expansions look like (see (\ref{3-090}), (\ref{5-060}) and (\ref{3-070})).
\begin{equation}\label{8-080}
u^{[1]}({\bf{x}})=\sum_{n=0}^{\infty}A_{n}^{[1]}J_{n}(k^{[1]}r)\cos(n\phi)~;~\forall\mathbf{x}\in\Omega_{1}~.~
\end{equation}
wherein:
\begin{equation}\label{8-090}
A_{n}^{[1]}=
A_{n}^{[0]}\left[\frac{2i\gamma^{[0]}}{\pi k^{[0]}h D_{n}}\right]~;~n=0,1,2,...~,~
\end{equation}
\begin{equation}\label{8-100}
A_{n}^{[0]}=2A^{i}\epsilon_{n}\exp(in\pi/2)\cos(n(\theta^{i}-\pi/2))~.~
\end{equation}
with the important reminder that $D_{n}$ {\it does not depend on} $\theta^{i}$.
If, as in panel (a) of fig.\ref{vary Nmax 2-4}, the field is restricted to the $n=0$ partial wave
\begin{equation}\label{8-110}
u^{[1]}({\bf{x}})=A_{0}^{[1]}J_{0}(k^{[1]}r)=
2A^{i}\left[\frac{2i\gamma^{[0]}}{\pi k^{[0]}h D_{0}}\right]J_{0}(k^{[1]}r)~;~\forall\mathbf{x}\in\Omega_{1}~,~
\end{equation}
I observe that this field does not depend on $\theta^{i}$. This result is true regardless of whether the choice $Nmax=0$ (as in fig.{\ref{vary Nmax 0-2}) is made or the frequency is low enough (as in fig.\ref{vary thetai}) for this to be justified, which fact explains remark 1) above.

As I increase the frequency, I must include more and more partial waves in the field representation, in which case I approach the situation described in fig.\ref{vary thetai}. Suppose that it is sufficient to include only the $n=1$ (in addition to the $n=0$) partial wave. Then
\begin{multline}\label{8-120}
u^{[1]}({\bf{x}})=A_{0}^{[1]}J_{0}(k^{[1]}r)+A_{1}^{[1]}J_{1}(k^{[1]}r)\cos\phi=\\
2A^{i}\left[\frac{2i\gamma^{[0]}}{\pi k^{[0]}h D_{0}}\right]J_{0}(k^{[1]}r)+4A^{i}i\cos(\theta^{i}-\pi/2)\left[\frac{2i\gamma^{[0]}}{\pi k^{[0]}h D_{1}}\right]J_{1}(k^{[1]}r)\cos\phi~;~\forall\mathbf{x}\in\Omega_{1}~.~
\end{multline}
and fig.\ref{vary thetai} should  translate graphically this basin response, at least at low frequencies. One immediately notes that the three panels (corresponding to three different incident angles) in this figure exhibit quite different low frequency response. Moreover, $\cos(\theta^{i}-\pi/2)=0$ when $\theta^{i}=0$ and $\cos(\theta^{i}-\pi/2)\neq 0$ when $\theta^{i}\neq 0$, which fact essentially explains remark 2) above.

Next, I consider the situation, in fig.\ref{vary thetai}, at the $f_{00}=0.776$Hz   resonance frequency. The resonance peak at this low frequency is hardly visible in panel (a) of the figure, but quite visible in the other two panels. Nevertheless, I am able to visually note the values of $I_{v}(f_{00})$ and find that they are equal to 0.355, 0.332 and 0.326 for $\theta^{i}=0^{\circ}$, $\theta^{i}=60^{\circ}$ and $\theta^{i}=80^{\circ}$ respectively. These resonant responses are quantitatively close, the small differences being due to the fact that even at resonance, the non-resonant partial waves bring a small contribution to the displacement field. Thus, I  conclude that the magnitude of the $f_{00}$ resonant response (as measured by $I_{v}$) does not depend on the incident angle. This conclusion is in agreement with the fact that, at this low resonance frequency, the displacement field essentially takes the form of (\ref{8-110}) which does not depend on $\theta^{i}$, knowing that $D_{0}$ does not depend on $\theta^{i}$.

I now return to fig.\ref{vary thetai} and examine the response at the second resonance frequency $f_{10}=1.25$Hz. I find that {\it no resonance peak is observable at this frequency in panel (a) (for $\theta^{i}=0^{\circ}$) whereas these peaks are observable in the two other panels (for $\theta^{i}\neq0^{\circ}$)}. A possible interpretation of this observation, which is  in the spirit of what is written in \cite{wd95}, is that the resonance frequencies vary with the incident angle since the  second well-formed maxima (marking the position of a resonance) in panels (b) and (c), which occur near $f=1.25$Hz, correspond to the first well-formed maximum that occurs near $f=1.70$Hz in panel (a) so that these maxima supposedly all correspond to the same resonance whose frequency of occurrence shifts with $\theta^{i}$. Theory, as embodied in (\ref{8-120}), tells a different, more plausible story, since it says that the second term in  (\ref{8-120}) vanishes for $\theta^{i}=0^{\circ}$ at all low frequencies in general, and at $f=f_{10}=1.25$Hz in particular, but does not vanish for $\theta^{i}=60^{\circ}$ and $\theta^{i}=80^{\circ}$, at which angles the amplitude of the resonance is particularly strong because of the near-vanishing value of $D_{1}$ at this (resonance) frequency $f_{10}$. Thus, once again, the locations (frequencies) of the resonances are governed
by the zeros of $D_{n}$ (i.e., and not by $\theta^{i}$; $D_{n}$ does not depend on $\theta^{i}$ either) whereas the presence, absence (more generally, amplitude} of a resonance is governed by the factor $A_{n}^{[0]}$ which depends explicitly on $\theta^{i}$ via the term $\cos(n(\theta^{i}-\pi/2))$.

Finally, I examine in fig.\ref{vary thetai} the response at the third resonance frequency $f_{01}=1.84$Hz. Once again, I note that the resonant peaks of $I_{v}$ are visible (and particularly strong for $\theta^{i}=0^{\circ}$) for all three incident angles. The theoretical reason for this lies once again in (\ref{8-120}) wherein the first term, which is non-zero for all incident angles, is large because $|D_{0}|$ is very close to zero (and the second term vanishes for this incident angle because of the $\cos$ factor and is otherwise small because $|D_{1}|$ is relatively-large at $f_{01}$ (due to the fact that this is not an eigenfrequency of the $n=1$ partial wave).

Thus, in resume, it can be said that the resonance frequencies do not depend on the incident angle, whereas the amplitude of the resonant response (here $I_{v}$, but, more generally, the displacement field in the basin) does depend on the incident angle. This is so because the response at an eigenfrequency depends not only on the corresponding eigenfunction, which  does not depend on the incident angle, but also on its participation factor $A_{n}^{[1]}$, which does depend on the incident angle (except for $n=0$) and can even vanish for certain $n$ and $\theta^{i}$.
\section{Examples, for several incident angles and increasing frequency, of non-resonant and resonant response, as is observable in the displacement field maps}\label{examples}
In the following figures I offer another way of visualizing both non-resonant and resonant basin response via the values taken by $|u^{[1]}(\mathbf{x})|$ throughout the domain occupied by the basin as well as by $|u^{[0]}(\mathbf{x})|$ in a portion of the bedrock (note however, that common practice is to present graphs of the displacement or velocity uniquely {\it on the ground}, assuming that the latter is free of settlement structures). Such displacement field {\it maps} have been published previously in \cite{sd00,sk05} relative to basins of quite different shape.

All the parameters I adopted in the previously-presented figures were also employed to produce the present field maps.
\subsection{Non-resonant response at $f=0.3$Hz}
An example of non-resonant response, occurring at the very low frequency $f=0.3$Hz, was already given in fig.\ref{softcircbasin7-8} in which it is observed that:\\ (a) $\max|u^{[1]}(r,\phi)|$ attains the value of $\approx 2.45$ (a.u.) regardless of the incident angle,\\(b) the field $|u^{[1]}(r,\phi)|$  within the basin is  concentrated around $r=0$, independent of $\phi$, and, slightly larger on the whole than the field $|u^{[0]}(r,\phi)|$ in the bedrock, \\(c) the map of displacement response within the basin does not depend on the incident angle (for reasons explained in the previous section), \\(d) the displacement field in the bedrock depends somewhat on the incident angle because the incident plus specularly-reflected fields $u^{i}+u^{r}$ are of amplitude comparable to that of the other component of the field in the bedrock, i.e. $u^{s}$, so that they can produce significant interference effects such as shadows whose spatial distribution varies with the incident angle, as is visible in the bedrock region of this figure.
\subsection{Resonant response at $f=f_{00}=0.776$Hz}
\begin{figure}[ht]
  \begin{center}
    \subfloat[]{
      \includegraphics[width=0.33\textwidth]{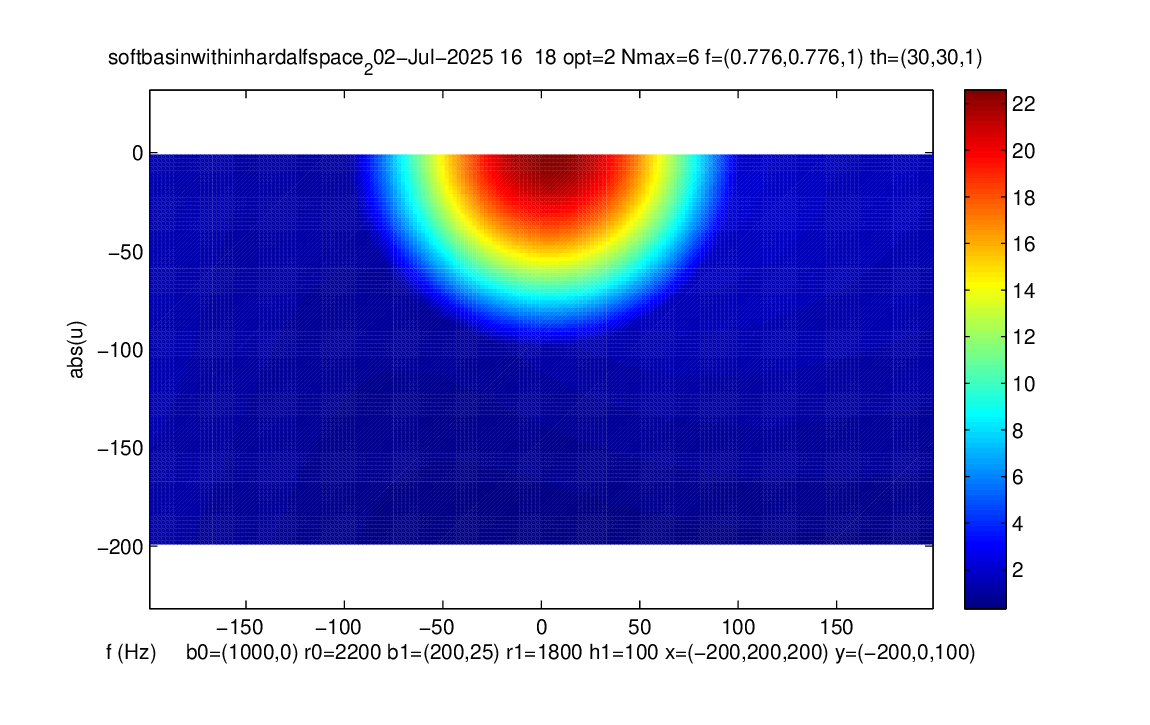}
      \label{a}
                         }
    \subfloat[]{
      \includegraphics[width=0.3\textwidth]{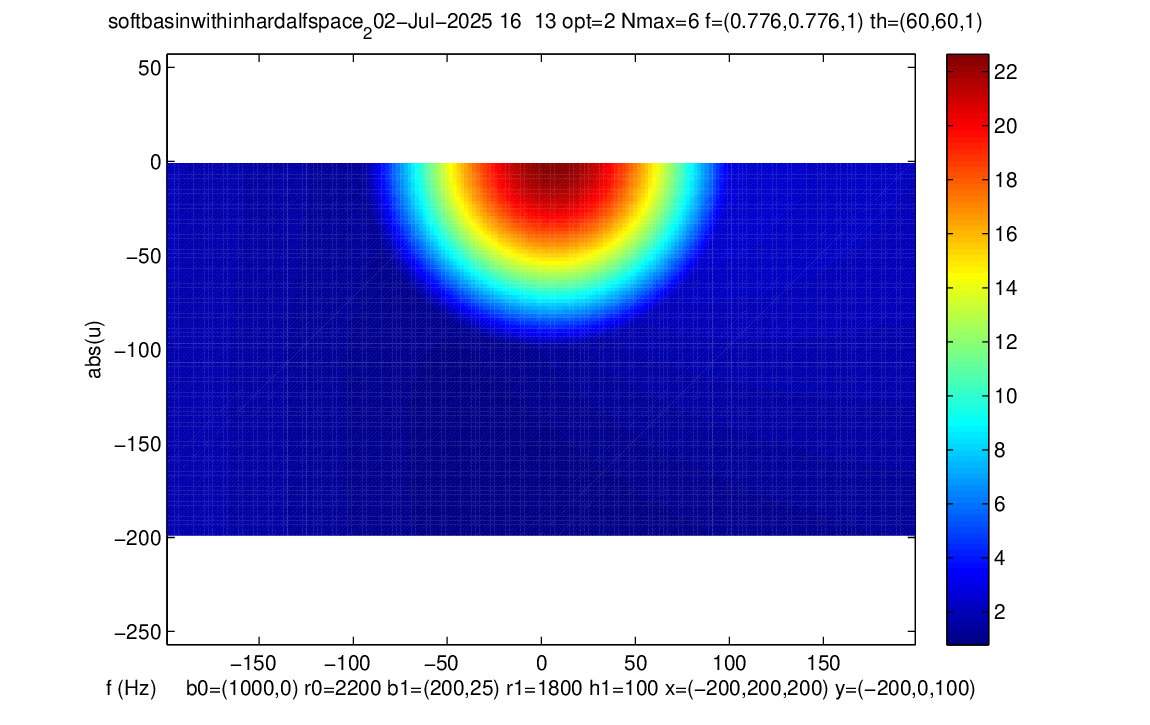}
      \label{b}
                         }
    \subfloat[]{
      \includegraphics[width=0.33\textwidth]{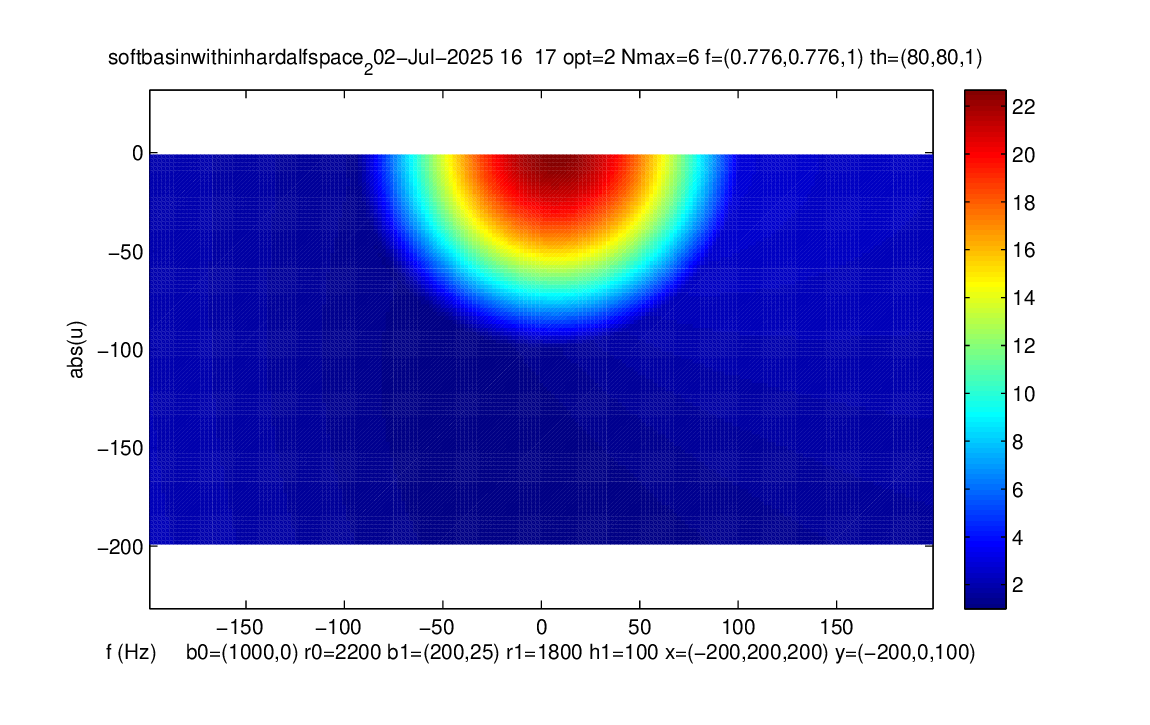}
      \label{b}
                         }

    \caption{Map of $|u^{j}(x,y)|; j=0,1$ for three different seismic loads at $f=0.776$Hz:\\
(a) $\theta^i=0^{\circ}$,  (b) $\theta^i=60^{\circ}$,  (c) $\theta^i=80^{\circ}$.
}
    \label{softcircbasin38-40}
  \end{center}
\end{figure}
 This figure applies to the first possible (in terms of ascending frequency) resonant situation. It is observed that: \\(a') $\max|u^{[1]}(r,\phi)|$ attains, nearly, and on the ground, the value of $\approx 22$ (a.u.) regardless of the incident angle, this value amounting to approximately ten times what was observed in the previous non-resonant situation; however the fact that $I_{v}$ attained only the modest value of 0.355 shows that there exists a poor correlation between $\max|u^{[1]}(r,\phi)|$ (near and on the ground) and the previous indicator of the amplitude of resonant response constituted by $I_{v}$, the reason for this being that the large displacement is concentrated  in a relatively- small portion of the basin (this is what I call a 'hotspot'), whereas for $I_{v}$ to be  large (i.e., close to 1) requires that the displacement be large throughout the basin,\\(b') the field $|u^{[1]}(r,\phi)|$ within the basin is independent of $\phi$, concentrated around $r=0$ (which explains the modest value of $I_{v}$), and, is much larger on the whole than the field $|u^{[0]}(r,\phi)|$ in the bedrock, \\(c') the displacement response within the basin does not depend on the incident angle; the reason for this is that at such a low frequency the basin field takes the single partial wave form of (\ref{8-110}) which does not depend on $\theta^{i}$), \\(d') the modulus of the displacement field in the bedrock seems to attain the largest value it would take in the absence of the basin, i.e., $\max|u^{i}+u^{r}|=2$,  which would indicate that $u^{s}$ is small even for small $r-h$, but the reason for this is not easy to discern from the SOV solution.
\clearpage
\subsection{Non-resonant response at $f=1.0$Hz}
\begin{figure}[ht]
  \begin{center}
    \subfloat[]{
      \includegraphics[width=0.3\textwidth]{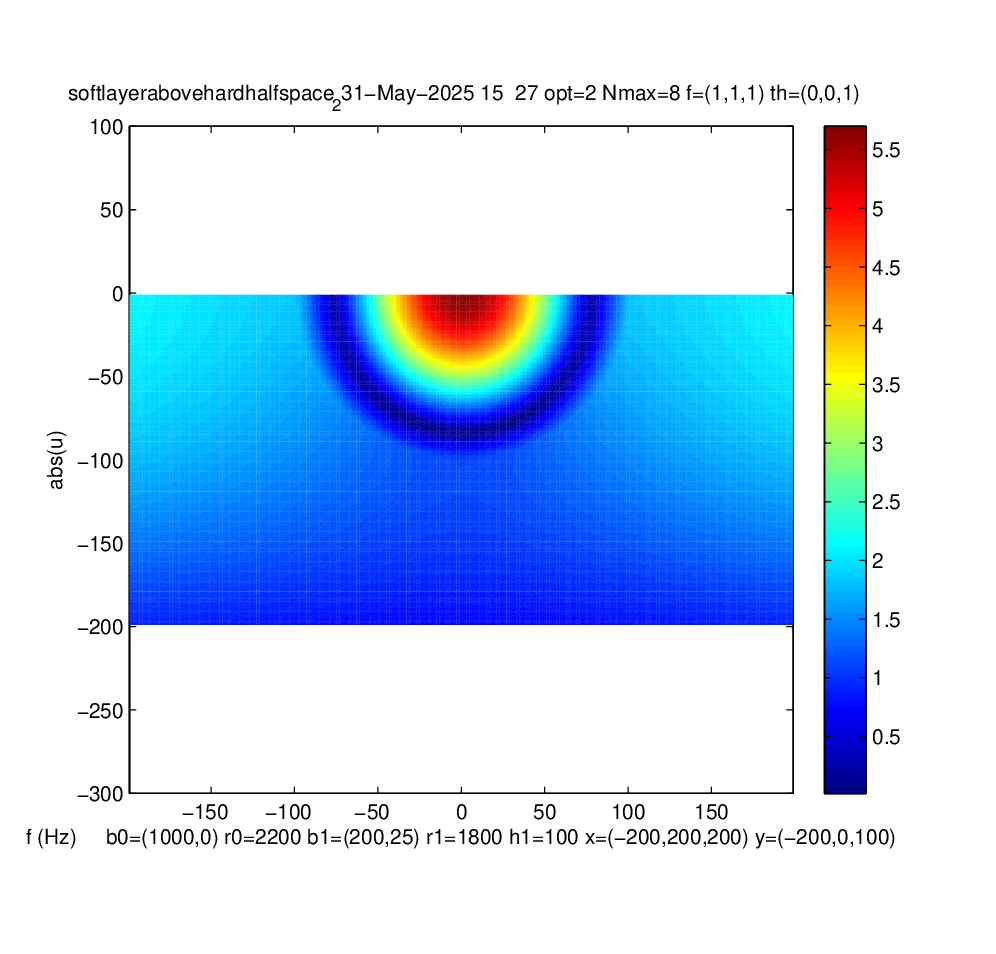}
      \label{a}
                         }
    \subfloat[]{
      \includegraphics[width=0.3\textwidth]{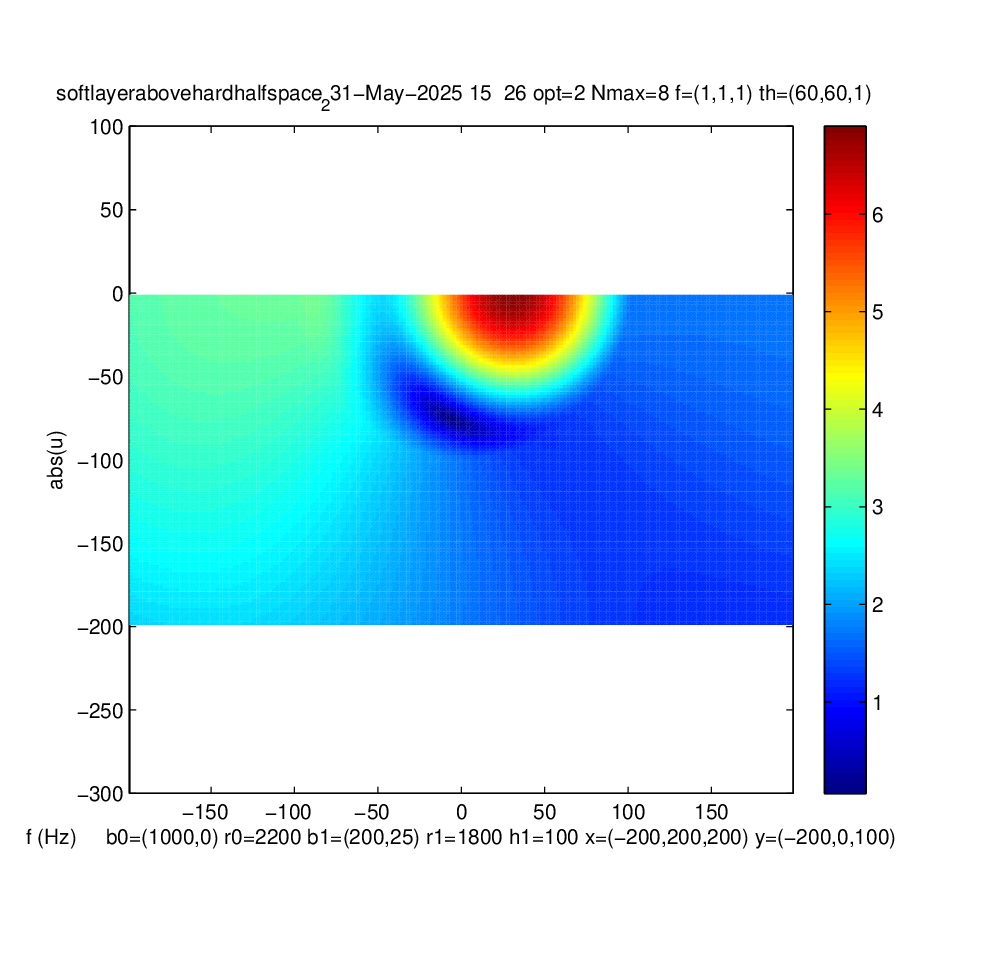}
      \label{b}
                         }
    \subfloat[]{
      \includegraphics[width=0.3\textwidth]{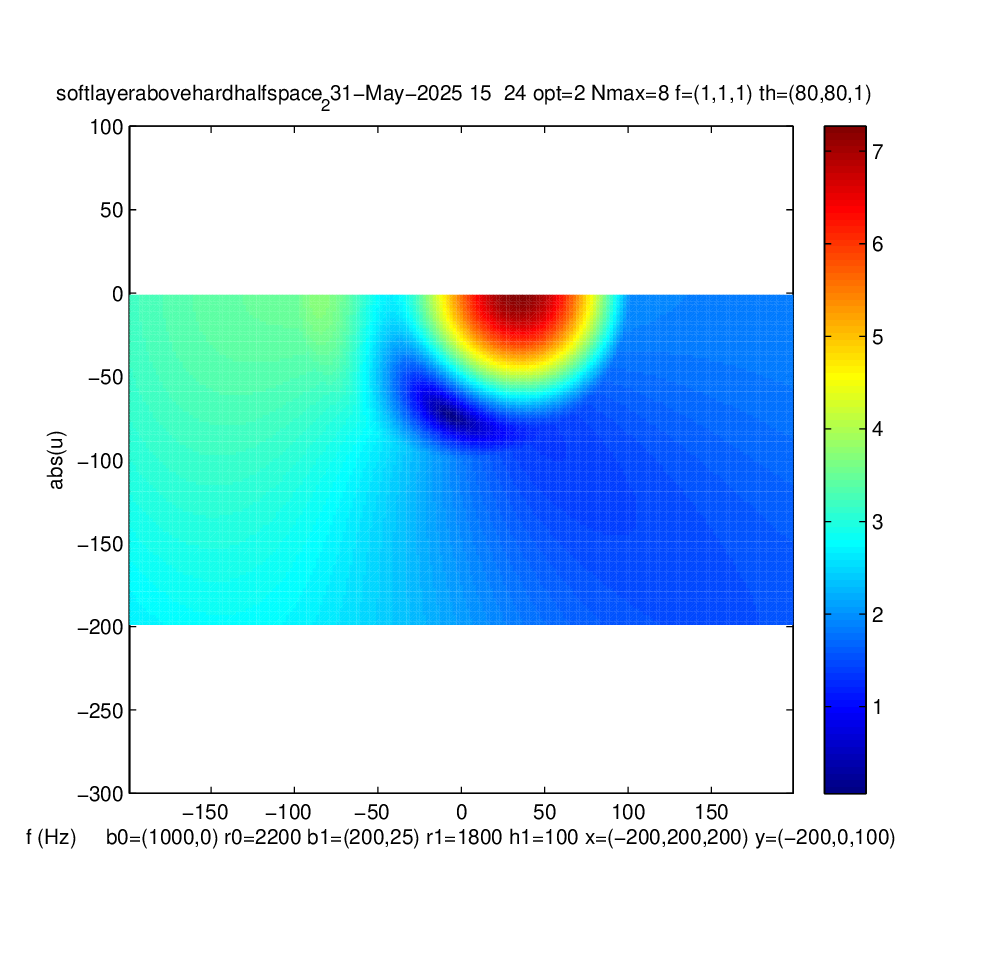}
      \label{b}
                         }

    \caption{Map of $|u^{j}(x,y)|; j=0,1$ for three different seismic loads at $f=1.0$Hz:\\
(a) $\theta^i=0^{\circ}$,  (b) $\theta^i=60^{\circ}$,  (c) $\theta^i=80^{\circ}$.
}
    \label{softcircbasin25-27}
  \end{center}
\end{figure}
The above figure is another example of non-resonant response, occurring, this time, at a low  frequency of $f=1.0$Hz. It is observed therein that: \\(e) $\max|u^{[1]}(r,\phi)|$ attains the (more than twice greater than at $f=0.3$Hz) value of $\approx 5.5-7.0$ (a.u.), now depending on the incident angle because the field requires the first two, instead of the former single partial wave(s) to be correctly-represented (and the second one depends on the incident angle, contrary to the first one), \\(f) the field $|u^{[1]}(r,\phi)|$ within the basin is again   concentrated around $r=0$, but now dependent on $\phi$ (since the second partial wave depends on $\phi$), and, somewhat larger on the whole than the field $|u^{[0]}(r,\phi)|$ in the bedrock, \\(g) the displacement field in the bedrock depends quite a bit  on the incident angle because the incident plus specularly-reflected fields $u^{i}+u^{r}$ are of amplitude comparable to that of the other component of the field in the bedrock, i.e. $u^{s}$, so that they can produce significant interference effects such as shadows whose spatial distribution varies with the incident angle, as is visible in the bedrock region of this figure, \\(h) for this reason, and the requirement of the second partial wave to correctly-represent $u^{s}$, the total field in the bedrock is observed to depend heavily on $\phi$ (as well as on $r$).
\subsection{Quasi-resonant response  at $f=1.77$Hz}

This frequency of $f=1.77$Hz is in between, and rather close to the two eigenfrequencies $f_{20}=1.70$Hz and $f_{01}=1.84$Hz.

\begin{figure}[ht]
  \begin{center}
    \subfloat[]{
      \includegraphics[width=0.3\textwidth]{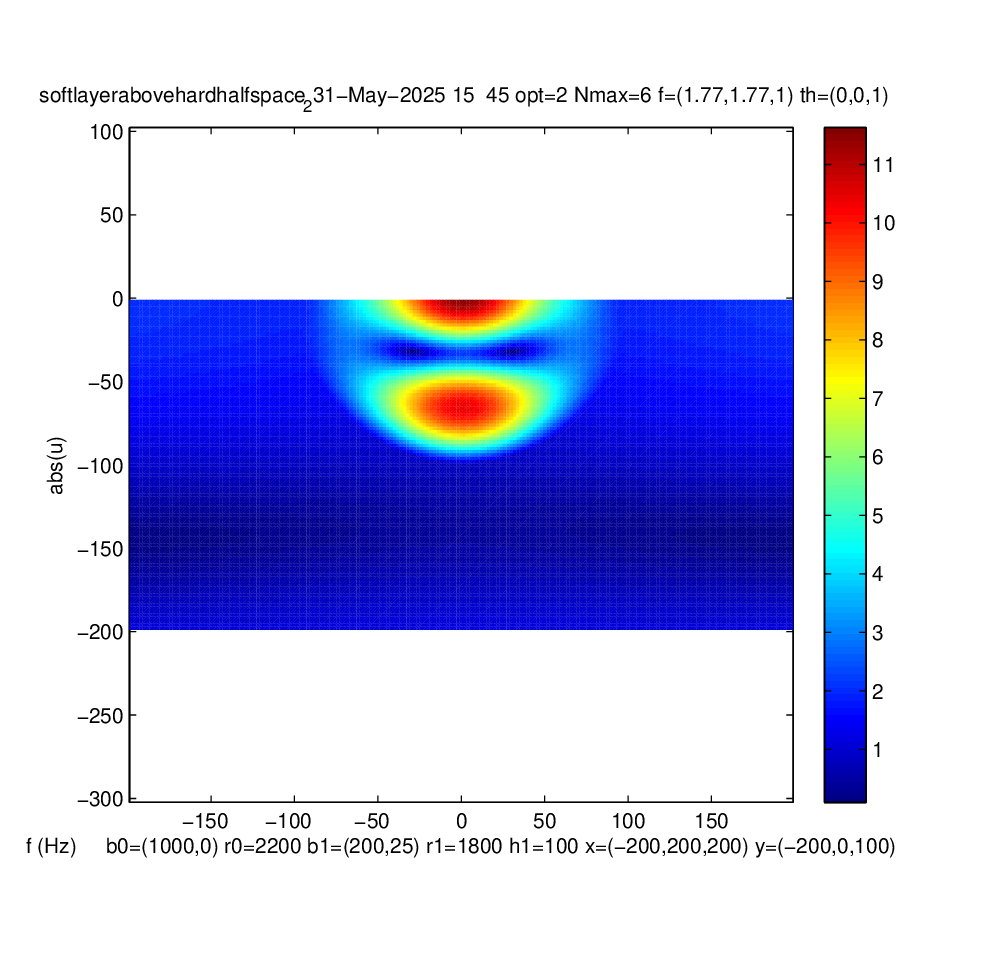}
      \label{a}
                         }
    \subfloat[]{
      \includegraphics[width=0.3\textwidth]{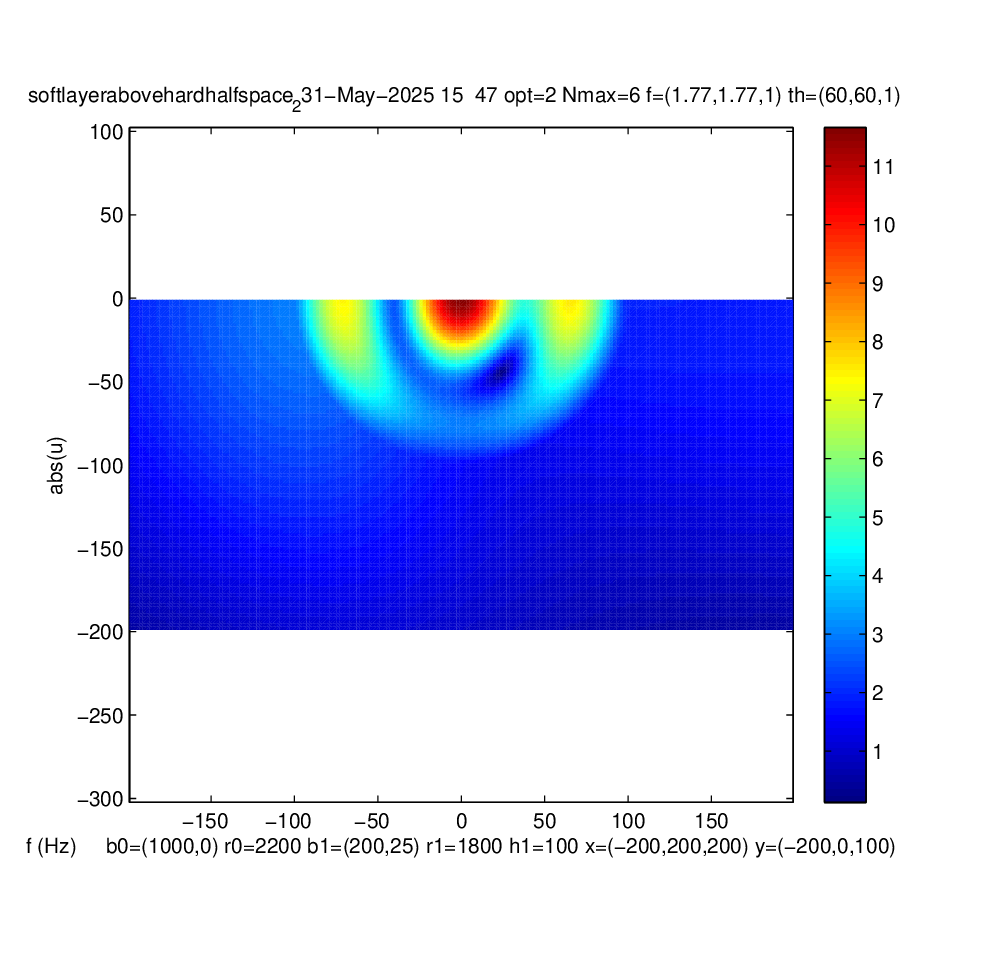}
      \label{b}
                         }
    \subfloat[]{
      \includegraphics[width=0.3\textwidth]{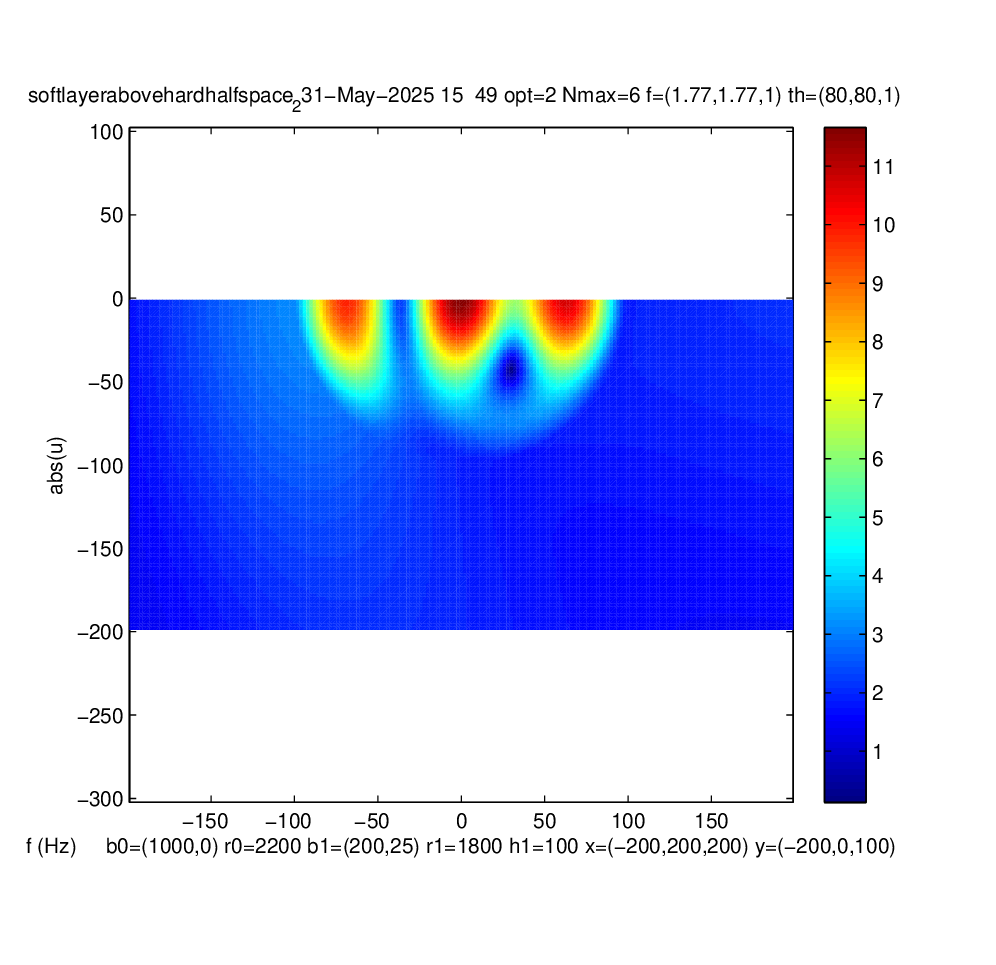}
      \label{b}
                         }

    \caption{Map of $|u^{j}(x,y)|; j=0,1$ for three different seismic loads at $f=1.77$Hz:\\
(a) $\theta^i=0^{\circ}$,  (b) $\theta^i=60^{\circ}$,  (c) $\theta^i=80^{\circ}$.
}
    \label{softcircbasin31-33}
  \end{center}
\end{figure}
At present, due to the closeness of the two eigenfrequencies, one can expect the response to be governed, both by the resonances at the latter two frequencies, and by non-resonant features. The fact that a $l=2$ eigenmode is involved here requires that three i.e., the $n=0,1,2$, partial waves  be taken into account to explain the spatial behavior of the displacement field at $f=1.77$Hz.

I observe the following features in fig.\ref{softcircbasin31-33}:\\
(a'') $\max|u^{[1]}(\mathbf{x})|$ attains the rather large value of $11.5$ for all three incident angles, but the spatial distribution of this basin displacement function is quite different from one angle of incidence to another, \\
(b'') since, at normal incidence, the displacement is large within two vertically-oriented  hotspots that occupy a large portion of the basin, it can be expected that the latter absorbs the most (not far from the entire portion) of the incident energy (flux) at this angle of incidence, this being substaniated by the results in fig.\ref{vary thetai} relative to $I_{v}$, \\
(c'') the presence of more than one hotspots (horizontally-oriented as soon as the incident angle departs from normal incidence) is probably due to the combined effect of (the necessary) three (rather than the previous two or one) partial waves in the field representations, \\(d'') it is observed that the horizontally-oriented hot spots occupy less of the region of the basin which means that the latter absorbs less energy (flux) at oblique incidence than at normal incidence, as is substantiated by the results in fig.\ref{vary thetai}, \\
(e'') the displacement field  in the bedrock is rather small, but heterogenous, and different from one angle of incidence to another.
\subsection{Resonant response  at $f=2.91$Hz.}
This frequency of $f=2.91$Hz is, for all practical purposes, equal to to the eigenfrequency $f_{02}=2.93$Hz, so that I now expect to see typical resonant response. However, as in the previous case, the chosen frequency is also  rather close to the  eigenfrequency  $f_{21}=2.81$Hz.
At present, due to the closeness of the two eigenfrequencies, one can expect the response to be governed, both by the resonances at the latter two frequencies, and by non-resonant features. The fact that an $l=2$ eigenmode is involved here requires that three i.e., the $n=0,1,2$, partial waves  be taken into account to account for the spatial behavior of the displacement field at $f=1.77$Hz.
\clearpage
\begin{figure}[ht]
  \begin{center}
    \subfloat[]{
      \includegraphics[width=0.3\textwidth]{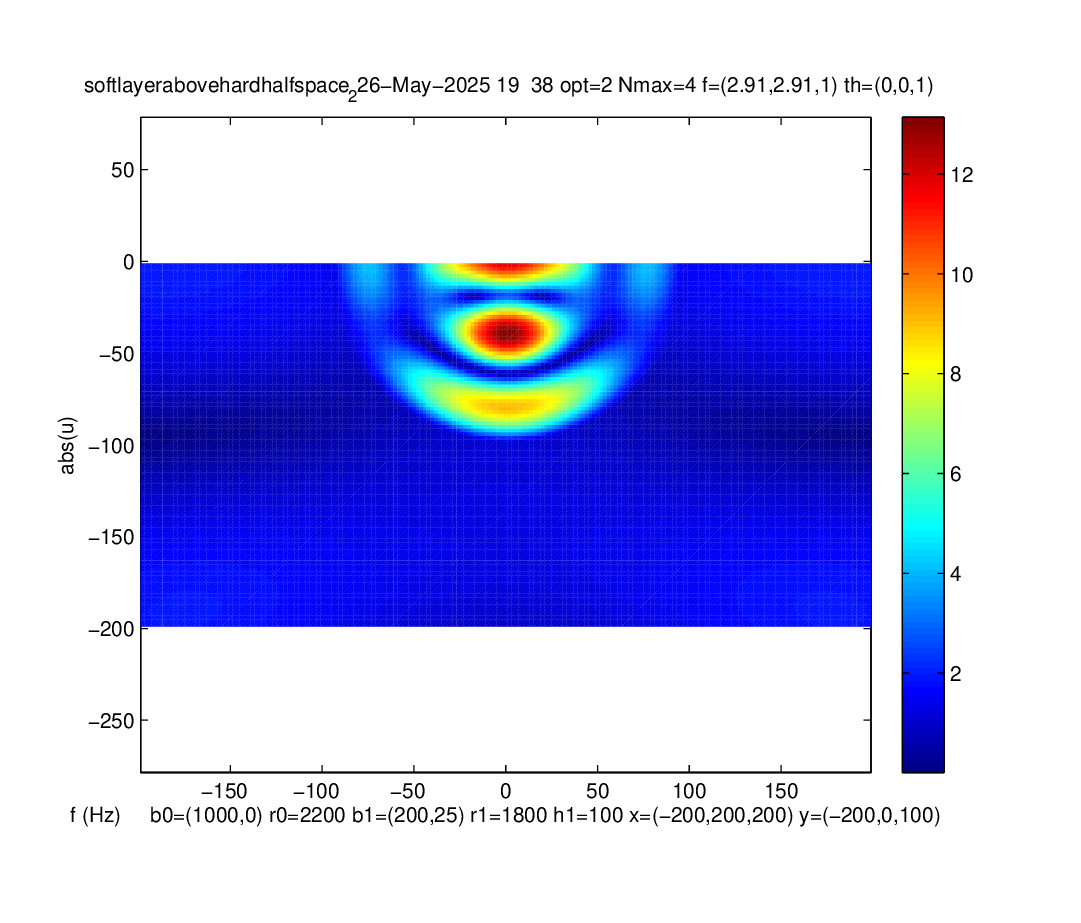}
      \label{a}
                         }
    \subfloat[]{
      \includegraphics[width=0.3\textwidth]{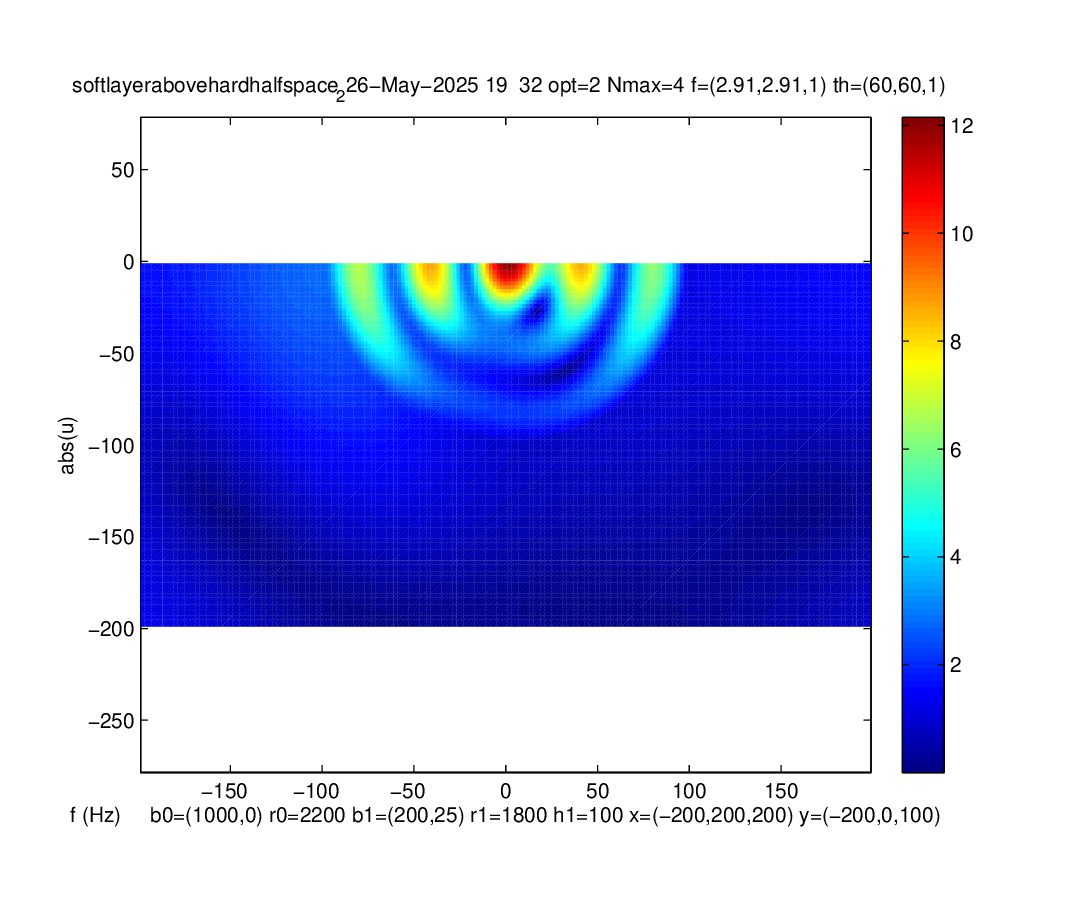}
      \label{b}
                         }
    \subfloat[]{
      \includegraphics[width=0.3\textwidth]{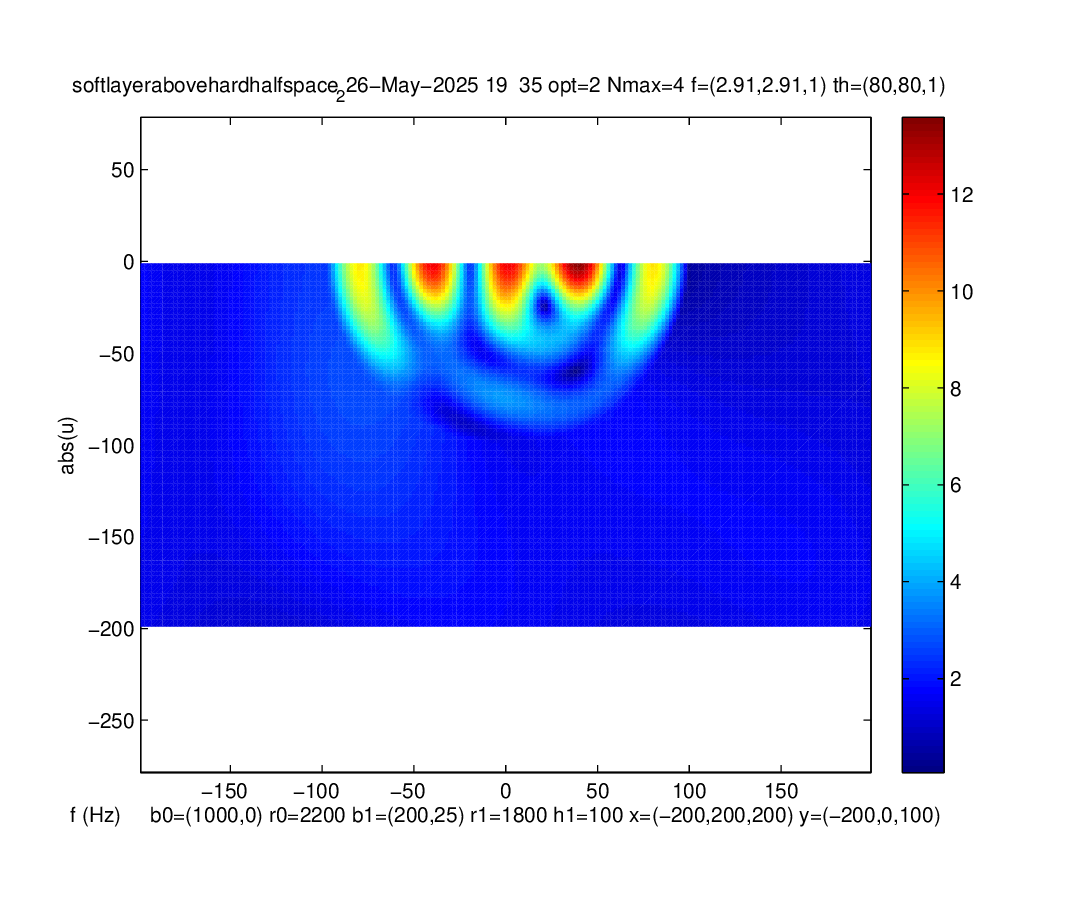}
      \label{b}
                         }

    \caption{Map of $|u^{j}(x,y)|; j=0,1$ for three different seismic loads at $f=2.91$Hz:\\
(a) $\theta^i=0^{\circ}$,  (b) $\theta^i=60^{\circ}$,  (c) $\theta^i=80^{\circ}$.
}
    \label{softcircbasin18-20}
  \end{center}
\end{figure}
I observe the following features in fig.\ref{softcircbasin18-20}: \\
(e''') $\max|u^{[1]}(\mathbf{x})|$ attains the rather large value of $12-13$ (a.u.) for the three incident angles, but the spatial distribution of this basin displacement function is quite different from one angle of incidence to another, \\
(f''') the displacement is large within  3 to 5  hotspots that occupy about half the basin, it can be expected that the latter absorbs about half of the incident energy (flux) for all three angles of incidence, this being substaniated by the results in fig.\ref{vary thetai} relative to $I_{v}$, (c'') the presence of more than one hotspots (horizontally-oriented as soon as the incident angle departs from normal incidence) is probably due to the combined effect of (the necessary) three  partial waves in the field representations, \\
(g''') the displacement field  in the bedrock is rather small, but heterogenous, and different from one angle of incidence to another.
\subsection{Quasi-resonant response at $f=4.48$Hz.}
\begin{figure}[ht].
  \begin{center}
    \subfloat[]{
      \includegraphics[width=0.5\textwidth]{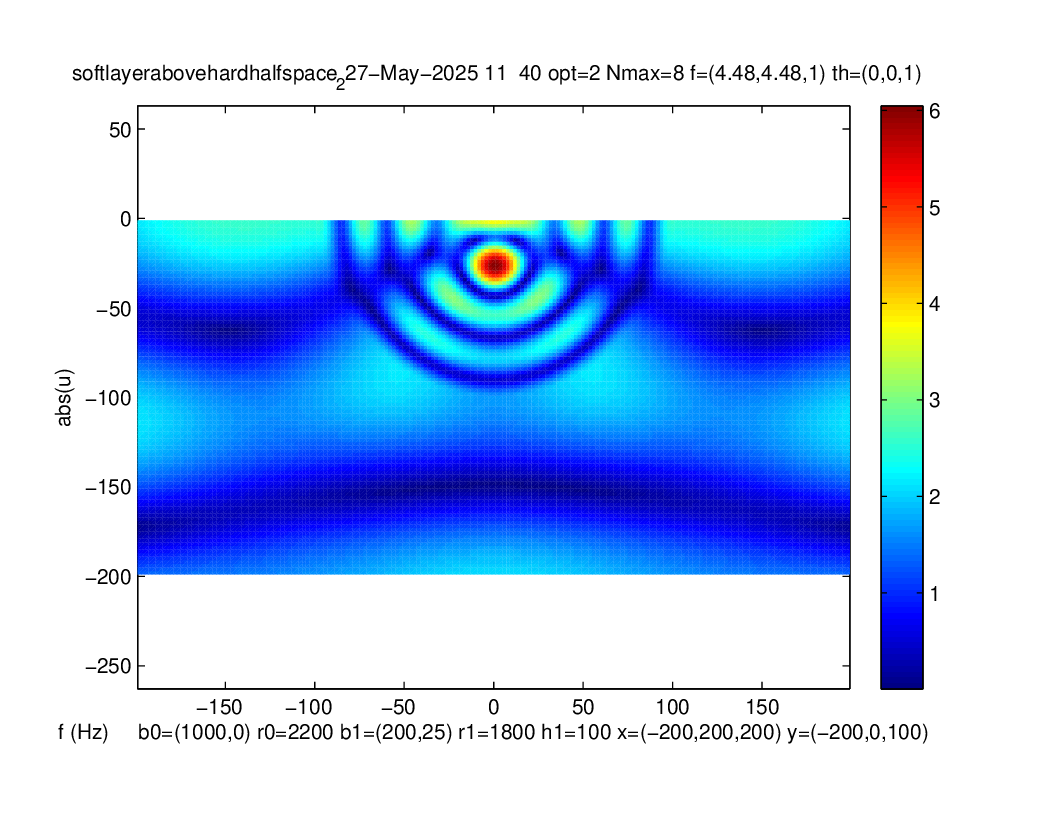}
      \label{a}
                         }
    \subfloat[]{
      \includegraphics[width=0.5\textwidth]{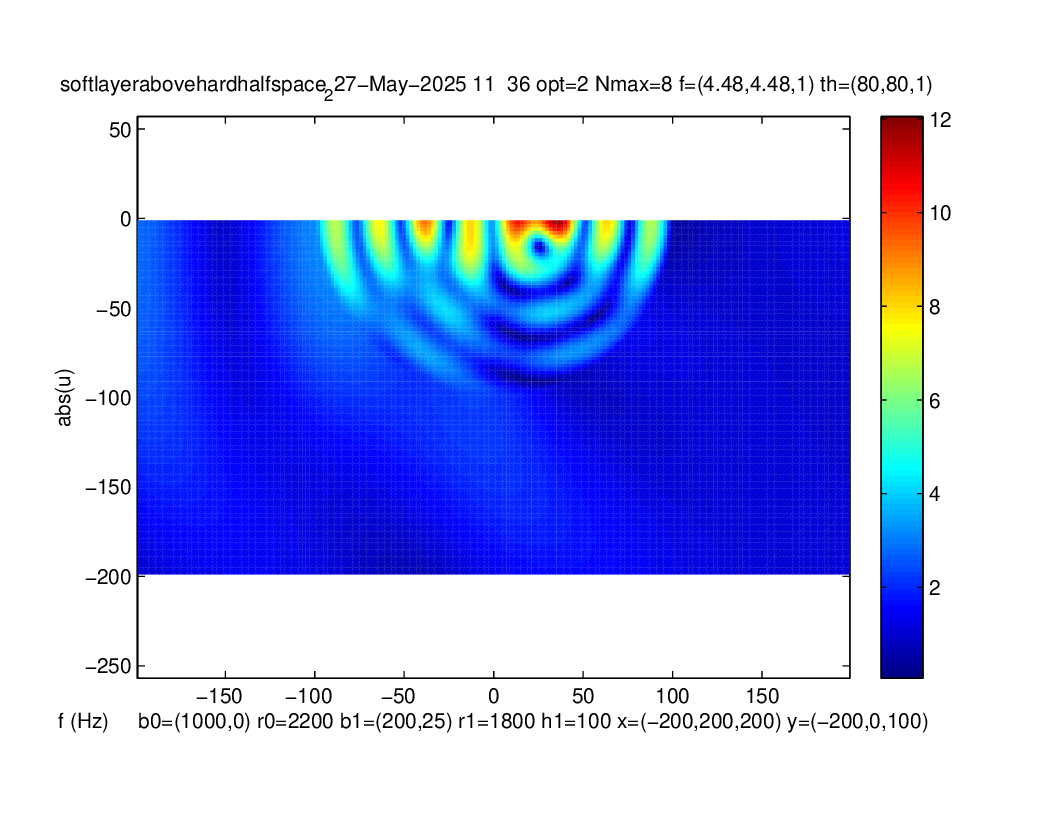}
      \label{b}
                         }
    \caption{Map of $|u^{j}(x,y)|; j=0,1$ for two different seismic loads at $f=4.48$Hz:\\
(a) $\theta^i=0^{\circ}$,  (b) $\theta^i=80^{\circ}$.
}
    \label{softcircbasin16+21}
  \end{center}
\end{figure}
At this high frequency, the computations require that $Nmax$ be increased to 8. From the theoretical point of view, the fact that $f=4.48$Hz is close to the eigenfrequency $f_{32}=4.39$Hz means that at least the four $n=0,1,2,3$ partial waves are necessary to account for the essential features of the response. As usual, this closeness of $f$ to the single $lm-$th eigenfrequency, leads us to expect that the response will be resonant in nature and due predominantly to the excitation of the $lm-$th eigenmode, in this case the $lm=32$ eigenmode, the latter being related to the $l=3$ partial wave. Fig.\ref{vary thetai} already alerts us to the fact that, at this eigenfrequency, $I_{v}$ is more than twice greater for the $80^{\circ}$ incidence than for normal incidence, so that it is interesting to find out if a similar effect is produced concerning the displacement field in the basin.

In fig.\ref{softcircbasin16+21} I plot the maps of this function for normal and $80^{\circ}$ incidence. It is observed that: \\
(e'''') $\max|u^{[1]}(\mathbf{x})|$ attains the fairly-modest value of 6 at normal incidence and 12 at $80^{\circ}$ incidence, these values being concentrated in (one at normal incidence, more than one otherwise) rather small hotspots,\\
(f'''') contrary to all that has been found until now, the most prominent hotspot in panel (a) (relative to normal incidence) is not in contact with the ground  so that the maximum of response is not at ground level and is rather modest, which fact could lead one to believe, if, as is usual, ground response is what is measured, that the basin is less 'dangerous' at this quasi-resonant frequency than at some non-resonant frequency, \\
(g'''') the reason for this unusual displacement response is as previously since the participation factor of the $l=3$ partial wave (and therefore of the $l=32$ eigenmode) vanishes at normal incidence because of the vanishing nature of the term $\cos(l(\theta^{i}-\pi/2))=\cos(l(0-\pi/2))$, \\
(h'''') the same term is practically equal to one when the incident angle is $80^{\circ}$ so that for this incidence the excitation of the $lm=32$ eigenmode is assured, which is what is observed in panel (b) of the present figure.
\section{Open questions}\label{disc}
%
\subsection{The spurious resonance question}
Many of the most interesting numerical results obtained by previous authors were the result of the employment of the Boundary Element Method (BEM) which is known \cite{wi22,cl25} to give rise to so-called spurious resonances for certain scattering configurations that should not produce resonant response. The fact that  publications such as \cite{sd00,sk05}, appealing to the BEM, contain maps of basin response that are manifestly of resonant nature at certain frequencies, could be an example of spurious resonances. However, in \cite{wi22}, I showed that in a case of a scatterer of canonical shape for which exact solutions can be obtained by separation of variables (also the case herein), the said SOV solution predicts either authentic resonances or no resonances at all. Thus, by comparison of my results herein for the basin having a semi-circular shape with those of \cite{sd00,sk05}, I take little risk in writing that their response maps, relative to a basin of more complicated shape than mine,  betray the existence of authentic resonances (although the authors of these publications do not use the term 'resonances' to qualify the cause of their hot spot phenomena).
\subsection{Are resonances the cause of significant amplification and response aggravation
for basins of arbitrary shape?}
On account of what I  wrote in the previous subsection, I am inclined to answer: yes. Moreover, amplification and response aggravation of manifestly resonant nature (but generally not qualified as such)  have been  documented  in many publications dealing with both measured and numerical seismic responses of soft basins of a great variety of shapes, embedded in hard half-spaces.
\subsection{Are surface waves the cause of significant amplification and response aggravation?}
Surface waves of the most well-known variety in seismology are the Love (L) and Rayleigh (R) waves \cite{ub73}. L and R waves seem to have been detected in the (both empirical and numerical) seismic responses of basins that are like layers of finite lateral extension. Moreover, it seems that the L and R waves originate at the corners of such basins \cite{bb80,sl25} at which locations the displacement is particularly strong. I have found no evidence of such large corner displacements, which fact leads me to doubt that surface waves are the cause of my amplified response results, but I know that in certain acoustical scattering configurations it is possible to establish an equivalence between surface wave generation and the occurrence of resonances \cite{rm10}.
\subsection{What is the most suitable variable to quantify the seismic response of soft basins?}
The terms 'amplification' and (related) 'response aggravation' originate in measurements and numerical simulations of seismic response at one or several points on the ground (assumed to be flat) covering a medium with both hard and soft components. There exist a large variety of definitions of 'amplification', the simplest of which is as follows. Imagine that the basin is absent so that the medium underneath the ground is entirely filled with hard rock. Then it it easy to show (for plane body wave excitation of amplitude 1 (a.u.)) that the maximum value of the modulus of the displacement field $|u|$ both within the underground $UG$ and on the ground $G$ is 2 (a.u.) so that any measurement of $|u|$  at point $P_{G}$ on the ground, underneath which there is some soft material embedded in the rock, that exceeds the value of 2 means that at $P_{G}$ the seismic response is amplified or equivalently, that $P_{G}$ is a location at which the seismic response is aggravated. To quantify such an amplification (designated hereafter by the symbol $\mathcal{A}$), one simply divides by 2 the measured or computed $|u|$. Such a definition extends without difficulty to points $P_{UG}$ in the underground.

Since $|u|$ depends not only on where $P_{G}$ is located on the ground but also on the characteristics of the incident seismic wave ($A^{i}$, $f$, and $\theta^{i}$ in our study), it is easy to conceive that potentially a large number of measurements, or numerical simulations thereof, are required to characterize a site such as the one under scrutiny herein in which the underground contains a soft basin.

For this reason, one often characterizes the site by replacing this  multitude either by an average or a maximum with respect to the location variables (on $G$ or in $UG$), or to one or some of the incident wave variables. Taking the maximum  was what I did in sect.\ref{examples} when, for instance, I qualified the enhanced seismic response within the underground as a whole by the single number 6 (3 after division by 2 to be in agreement with the aforementioned simple definition of amplification) for the fixed values of $A^{i}$, $f$, and $\theta^{i}$. As one could  see in this, and the other examples, the possibility exists that $|u|$ takes on even much larger values, which fact suggests that {\it there might not be any limit to the values taken by $\mathcal{A}$}, which is a rather annoying feature for a number that is chosen to characterize a physical, measurable, variable (see, for instance, \cite{gr05,sd00} for other examples of site characterization in which computed amplifications are found to be very large, for instance 16 in \cite{sd00}).

This feature is aggravated by another one: when attempting to compare the $\mathcal{A}$'s of two sites, via one publication or another in which the definition of $\mathcal{A}$ is uncertain and/or changing, it becomes impossible to draw meaningful conclusions about the amplifications of these sites. This, of course, is particularly true for amplifications that derive from actual measurements (because of the differences of measurement procedures), but it is also a feature of amplifications that derive from numerical computations (because each of these is subject to certain limitations (for instance, herein, it is important to choose and specify the minimal number of partial waves that are necessary to achieve meaningful predictions of the response of the basin) and errors that are not always known or specified in the publications).

The problem of the possibly-infinite (in any case, large) values of $\mathcal{A}$ attracted the attention of the authors of \cite{sd00} since they recognized that it was the result of computations for a basin filled with a soft {\it lossless} medium. They took into account the damping of the medium in an apparently-simple manner. However, one gets the feeling that the concept of amplification, as it applies exclusively to the ground measurements or predictions, cannot tell the full story of what is going on below the ground,
 notably when the medium in the underground is lossy. What I am trying to say is that the
 reduction of response due to damping in the underground is an affair that necessarily involves not only a material damping coefficient such as $Q$ (in addition to the other constitutive parameters of the basin and bedrock), but also the  geometrical aspects (notably the volume) of the basin as a whole, and it is difficult to perceive how such complexities are reflected in a quantity such as $\mathcal{A}$, all the more so than the authors of publications dealing with the prediction of $\mathcal{A}$ in lossy environments usually do not appeal to the necessary theoretical (as opposed to numerical) tool to account for the subtleties of site response in lossy (and even lossless) environments.

 To fully appreciate the magnitude of this problem, I return to the case in which the site involves no lossy material. If, as is usually assumed, the underground is of semi-infinite extent in the vertical sense and infinite in the horizontal sense, and, in addition, no substantial approximations are made in the theoretical model, then it is found (or should be found) that the response of the basin configuration, as represented by $\mathcal{A}$,  is finite (notwithstanding that it can be large at certain frequencies), which fact implies that there exists a phenomenon of damping in the scattering configuration that has nothing to do with material damping. As is now rather well-known, this phenomenon is {\it radiation damping} \cite{mw94}, the meaning of which is that the scattering process produces a wavefield that is outgoing from the scatterer (here, the basin) towards the outer confines of the underground half-space. This outgoing wave necessarily carries part of the incident energy away from the basin towards outer space where it is, for all practical purposes, lost. This loss is radiation damping.

  When, in addition, the scatterer is lossy, two (rather than the previous one) types of damping come into play: radiation damping (RD) and material damping (MD), and it is legitimate to inquire as to what fraction of a quantity such as $\mathcal{A}$ should be attributed to RD and what part to MD, but as far as I know, there is nothing in the concept of amplification that enables this evaluation.

 This is why (in addition to reasons alluded-to previously) I undertook the conservation of energy (flux) analysis in the present study. However, lest I be misunderstood: the adoption of the loss functions that appear in my conservation law are  not meant to replace $\mathcal{A}$, all the more so than this latter quantity is close, or identical to, what can, and is usually, measured, whereas the volumic loss function that appears in my conservation law would require measurements of displacement at an enormous number of points within the basin (assuming that the bedrock is lossless) which is impossible from a practical viewpoint. However, it is possible in a computational context (provided one disposes of a decent laptop computer, as is commonplace today. The secondary advantage of a concept such as the volumic loss function is that it provides a single number to quantify the response for a given scattering configuration (the descriptors of which are the scatterer, its environment, the frequency and other characteristics of the seismic source or incident wave, in contrast to $\mathcal{A}$ which, for the same given descriptors is usually required to be measured or computed at a host of points on the ground).
 \section{Conclusion}
 After this lengthy discussion, a short conclusion. The theoretical apparatus leading to a conservation of energy (flux) law and a rigorous, complete and explicit solution of the scattering problem was primarily enabled by the simplicity of the (canonical, cylindrical semi-circular) model of the basin and the SH polarization of the incident wave. I am fully aware that this simplicity obviates a full understanding of what occurs when a real-life seismic wave hits a real-life basin, but the introduction of complexity, such as that of more general basin shapes (see, e.g., \cite{sd00,sk05}) leads to seismic responses that are quite similar to the ones presented herein, so that it is legitimate to consider  that my results reflect quite well what happens in more realistic configurations.

 Contrary to what is contained in practically all the previous, essentially numerical, publications treating this subject, herein I give a full explanation of what is going on when a seismic wave propagating in a hard half space strikes a soft basin. The key word to describe the response is {\it resonant}. I am not the first author to have noted that the seismic response (which translates to either $|u|$ or $\mathcal{A}$ versus $f$) of basins is generally marked by a succession of peaks. Usually, the first peak (in terms of ascending $f$) is what most interested the other authors, which is why they named it the 'fundamental resonance peak', without explaining what they meant by 'fundamental resonance', nor why the other peaks did not merit being associated with the term 'resonance'. This feature can be explained by the essentially-numerical nature of most of these previous studies, but it is nevertheless surprising that practically all the ingredients of resonant basin response were available since the abundantly-cited publication of Trifunac in 1971 \cite{tr71}. I say 'practically all' because  even Trifunac did not employ the term 'resonances', nor therefore explain how it becomes the principal agent of basin amplified response. As far as I know, my 1995  publication \cite{wi95} is the first one to provide these explanations. Since then, some disarming statements have been made such as: 'the resonance frequencies depend on the incident angle' \cite{wd95,sd00}, which are in total contradiction with what was written in \cite{wi95} and proved once again herein.

 This is the principal reason for having undertaken the present study. The secondary reason has, of course, to do with the proposal of using conservation of energy (flux) to qualify more completely the notions of (site) amplification and aggravation, by showing that material loss $L_{v}$ (in the volume of the basin)  is not the only loss mechanism, but that radiation damping is another one ($L_{r}$) and that the addition of these two is practically equal to the energy (flux) carried by the incident seismic wave. This feature of the present study was not just a semantic exercise, since I also provided the theoretical and numerical means (even for basins of general shape and dispersive, complex, dispersive shear modulii) to represent and compute each of these two loss components as well as verify (for the basin of semi-cylindrical shape) that they are indeed such as to verify the conservation of energy (law).

 The next task, not undertaken herein, is to reinstate the above-ground built component of a typical settlement, and show that the typically-resonant phenomena that come into play for above-ground structures are essentially the same as those for below-ground structures (such as basins) so that at resonance frequencies the above-ground structure becomes another (in addition to the basin) pit  of absorbed energy responsible for its damage or destruction (introductions to this subject can be found in \cite{gr05,wi16}).

\end{document}